\documentclass[aps,prd,twocolumn,nofootinbib,amsmath,amssymb,showpacs,superscriptaddress]{revtex4-1}

\newcounter{intro}

\usepackage{graphics}
\usepackage{graphicx}
\usepackage[usenames,dvipsnames]{color}
\usepackage{xcolor}
\usepackage{hyperref}
\hypersetup{
    colorlinks=true,
        linkcolor={blue!80!black},
        citecolor={blue!80!black},
        urlcolor={blue!80!black} 
}
\usepackage{mwe}
\usepackage[subnum]{cases}
\usepackage{lipsum}
\graphicspath{{./figs/}}

\begin{document}

\title{How the QCD trace anomaly behaves at the core of twin stars?}

\author{Jos\'e C. {\sc Jim{\'e}nez}}
\affiliation{Departament of Astrophysics, Brazilian Center for Research in Physics (CBPF),\\ Rua Dr. Xavier Sigaud, 150, URCA, Rio de Janeiro CEP 22210-180, RJ, Brazil}
\affiliation{Universidad Tecnológica del Perú, Arequipa - Perú}

\author{Lucas {\sc Lazzari}}
\affiliation{Institute of Physics and Mathematics, Federal University of Pelotas, \\
  Postal Code 354,  96010-900, Pelotas, RS, Brazil}

\author{Victor P. {\sc Gon\c{c}alves}}
\affiliation{Institute of Physics and Mathematics, Federal University of Pelotas, \\
  Postal Code 354,  96010-900, Pelotas, RS, Brazil}
\affiliation{Institute of Modern Physics, Chinese Academy of Sciences,
  Lanzhou 730000, China}

\date{\today}


\begin{abstract}
  We investigate the behavior of the dense and cold (normalized) QCD trace anomaly, $\Delta$, in the interior of twin neutron stars (obtained from several sets of equations of state in agreement with modern compact-star and multimessenger data) satisfying static and dynamic stability conditions. We scan {the formed twin-star} parameter space in order to look for effects caused by the presence of a strong first-order phase transition connecting hadron and quark phases by means of a Maxwell construction. We found robustly that $\Delta$ suffers an abrupt decrease around the transition point, even reaching large negative values ($\Delta\simeq-0.35$), in marked contrast to current studies pointing out a smooth behavior with $\Delta\gtrsim 0$ at all densities. Besides, we characterize the behavior of {the recently defined} conformal factor, $d_{c}$, in {our four categories of} twin stars for which we perform comparisons with theoretical constraints {put}, e.g. {by} Bayesian studies {sometimes} adjusted to agree with {perturbative} QCD {at high densities}. {These analyses} allow us to hypothesize modifications in the strong QCD coupling in dense nuclear matter with a strong thermodynamic discontinuity.
\end{abstract}

\maketitle

\section{Introduction}

According to the Standard Model of Particle Physics, every observable concerning strong nuclear interactions should be derivable from Quantum Chromodynamics (QCD) having two main (poorly understood) aspects: color confinement and chiral symmetry breaking ($\chi$SB) \cite{Ioffe:2010}. In order to probe them, one would need a reliable knowledge (at some fiducial scale) of the strong coupling $\alpha_{s}$ and the bare quark masses for the active flavors \cite{Ioffe:2010}. Currently, we have only a theoretical control on QCD at high energies due to {\it asymptotic freedom}, i.e. $\alpha_{s}\ll 1$, thus allowing us to characterize it as a chirally symmetric and deconfined QCD phase, altogether called perturbative QCD (pQCD) \cite{Ioffe:2010}. Nevertheless, at hadronic energies, the QCD coupling increases to a non-small value ($\alpha_{s} \sim 1$), thus making extremely difficult studying non-perturbative $\chi$SB and (de)confinement \cite{Ioffe:2010}.

Although there are remarkable advancements towards probing these non-perturbative features of QCD vacuum in collider physics, e.g. the LHC at CERN (see, e.g., Ref. \cite{Gehrmann:2021qex}), its emergent thermodynamic observables at equilibrium\footnote{The non-equilibrium situation concerning transport properties is even far more complex. See, e.g. Ref. \cite{Barata:2022utc} and references therein.} to probe relevant regimes, e.g. its equation of state (EoS), are still not well understood \cite{Kapusta:2006pm}. For instance, lattice QCD (LQCD) only provides reliable results for hot QCD matter \cite{Bazavov:2009zn,Borsanyi:2013bia} with background magnetic \cite{Bali:2011qj,Brandt:2023dir} and electric fields \cite{Endrodi:2023wwf} or finite isospin densities \cite{Brandt:2022hwy,Abbott:2023coj}, all of them being of relevance for heavy-ion physics. Nevertheless, most strongly interacting matter in the Universe is baryonic, being the extreme density situation present at the core of heavy neutron stars (NS), where a deconfinement transition is possible at $T=0$ \cite{Glendenning:2000}. Unfortunately, the LQCD method to finite baryon densities is strictly forbidden by the {fermionic} sign problem \cite{Nagata:2021ugx}.

To alleviate this difficulty at finite baryochemical potentials, $\mu_{B}$, available NS data (related to their mass-radius ($M$--$R$) diagram \cite{Fonseca:2016tux,Linares:2018ppq,Miller:2019cac,Riley:2019yda,Cromartie:2019kug} and tidal deformabilities from binary NS mergers \cite{LIGOScientific:2017vwq}) is currently employed to constrain the behavior of the NS EoS at intermediate (non-perturbative) densities (see, e.g., \cite{Annala:2021gom}). Broadly, their approach consists in interpolating theoretical results between the extreme cases of low and high $\mu_{B}$, i.e. chiral effective theory (CET) \cite{Hebeler:2013nza} and pQCD \cite{Kurkela:2009gj}, respectively. After that, the outcomes are adjusted to respect astrophysics and experimental nuclear data, thermodynamic consistency and causality limits \cite{Sorensen:2023zkk}.

While one might be tempted to believe that the physics of (de)confinement and $\chi$SB are hidden within these continuous interpolations (as occurs when using heavy-ion data to constrain the baryon-free QCD EoS at $T\neq0$ remarkably coinciding in most cases with the corresponding LQCD results \cite{Ratti:2018ksb}), most of these studies mainly favor smooth EoSs at all densities (e.g. see also Ref. \cite{Somasundaram:2021clp}). {Most of these works} also point out that the (adiabatic) speed of sound, $c^{2}_{s}{~\equiv~}dP/d\epsilon$ (being `$\epsilon$' the energy density and `$P$' the pressure), of dense and cold QCD matter surpasses the conformal bound, $c^2_{s}=1/3$, in the non-perturbative sector of baryon densities, even reaching values near the causality limit, $c^2_{s}=1$ \cite{Tews:2018kmu}, something which is in tension with field theory studies \cite{Cherman:2009tw,Bedaque:2014sqa}. Furthermore, Ref. \cite{Annala:2019puf} even suggest the presence of quark matter (QM) in some maximal-mass NSs if $c^{2}_{s}\gtrsim 0.5$. {Related to this}, it should be noticed that the above findings have received support from agnostic Bayesian studies \cite{Somasundaram:2022ztm,Komoltsev:2023zor}.
 
\begin{figure}[!t]
  \vspace*{-0.65cm}	
  \hspace*{-0.6cm}
  \includegraphics[width=.56\textwidth]{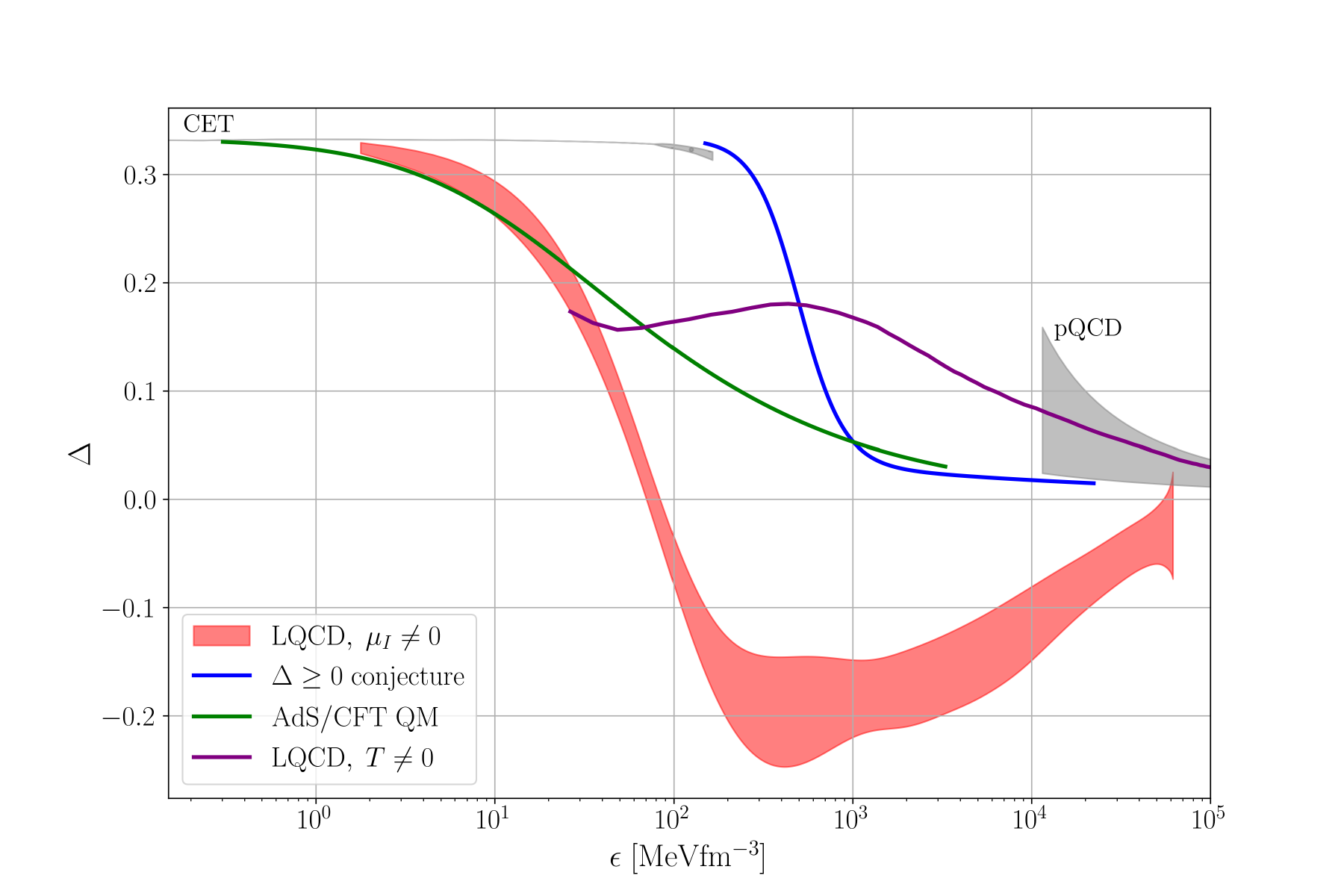}
  \caption{\label{fig:others}Behavior of the normalized in-medium trace anomaly vs energy density ($\Delta$ vs $\epsilon$) for QCD matter in different extreme conditions. In particular, we display known findings for i) the conjectured positiveness of Ref. \cite{Fujimoto:2022ohj} based upon the maximum mass bound of NSs ($\Delta \geq 0$ conjecture), ii) AdS/CFT QM~\cite{Hoyos:2016zke} (see Appendix A for details), iii) baryon-free LQCD at finite temperatures (LQCD, $T\neq 0$) \cite{Bazavov:2009zn,Borsanyi:2013bia} and iv) LQCD for cold isospin matter (LQCD, $\mu_I\neq 0$) ~\cite{Abbott:2023coj}.  {Notice that} we superimposed {these physically different $\Delta$'s} in the same plot to clearly {exhibit} their sign differences. Besides, the bands CET/pQCD are only given for the case with $\mu_B \neq 0$ {in $\mu_{B}$-regions where both are valid and reliable \cite{Fujimoto:2022ohj}}.}
\end{figure}

In this sense, along the last few years, great attention was given to get insights for this large values of $c^{2}_{s}$ and verify if it is a realistic property of dense matter or only an artificial effect required to make sense of the employed (agnostic or not) interpolations \cite{Brandes:2022nxa}. Among these investigations, the work of Ref. \cite{Fujimoto:2022ohj} had taken a different path by connecting $c^{2}_{s}$ to the so-called (normalized) dense QCD {\it trace anomaly} $\Delta \sim \epsilon - 3P$ (to be fully defined in Sec. \ref{sec:Tanomaly}) which, broadly, is a measure of the quantum breaking of scale-conformal invariance\footnote{It is also known as the `QCD interaction measure' at finite temperatures. As we will see below, this is not appropriate since it assumes its positiveness which, in general, is not always the case.}. {In particular, the authors of Ref. \cite{Fujimoto:2022ohj}} (based on modern NS data) {along with} other past studies (from e.g. baryon-free LQCD \cite{Bazavov:2009zn,Borsanyi:2013bia} at all temperatures, pQCD with CET \cite{Fujimoto:2022ohj} at their range of validity, and holographic AdS/CFT results \cite{Hoyos:2016zke}) {went a step further by conjecturing the numerical positivity of} $\Delta$ at all densities. On the other hand, isospin LQCD found $\Delta<0$ for a large range of intermediate densities, i.e.  {within} its non-perturbative region \cite{Brandt:2022hwy,Abbott:2023coj}. In fact, Ref. \cite{Abbott:2023coj} indicate that their large negative $\Delta$ is mostly caused by a pairing term (akin to the CFL phase of QM). We display the behavior of these cited results in Fig. \ref{fig:others}.

Thus, from the pure definition of $\Delta$, there is no general criterium (like a non-perturbative QCD theorem) preventing us from getting a negative behavior at intermediate $\mu_{B}$ since it depends on the interactions introduced in the Lagrangian defining the Landau thermodynamic potential \cite{Kapusta:2006pm}. For instance, it can be proven \cite{Landau:1987} that for pure electromagnetic interactions, one has $\Delta\geq{0}$. However, it has also been proven within the NJL and leading-order CEF \cite{Lu:2019diy} that at $T=0$ and finite isospin densities, a transition from $\Delta\geq{0}$ to $\Delta<{0}$ occurs, possibly corresponding to a BEC-BCS crossover. Besides, a negative $\Delta$ at $T\neq{0}$ in the primordial universe for pion condensation \cite{Vovchenko:2020crk} had been found, being correlated to both large $c^{2}_{s}$ and leptonic densities.
 
{In the case of the quark-gluon plasma, i.e. hot QCD matter, (formed at early stages after ultrarelativistic} heavy ion collisions occur), {one has that} $\Delta \rightarrow 0$ as hot pQCD reaches its conformal limit at very high temperatures. However, at intermediate $T$, {LQCD indicates that it is positive definite. Unfortunately,} there is no clear explanation for this phenomenon within QCD {or effective versions}. In fact, up to now, there are few {studies favoring} this trend, e.g. Ref. \cite{Albright:2015fpa}. {Interestingly}, this issue was also studied in Ref. \cite{Kharzeev:2007wb} from a hydrodynamic perspective relating the QCD bulk viscosity, $\zeta_{\rm QCD}$, and {the corresponding QCD} trace anomaly {obtained from LQCD}, i.e. $\zeta_{\rm QCD}\sim (\epsilon-3P)_{\rm LQCD}$, stating that it should always be positive\footnote{Explicitly \cite{Kharzeev:2007wb}
\begin{equation*}
    \zeta_{\rm QCD}=\frac{1}{\omega_{0}}\left\lbrace{T^{5}\frac{\partial}{\partial{T}}\frac{(\epsilon-3P)_{\rm LQCD}}{T^{4}}+16|\epsilon_{v}|}\right\rbrace,
\end{equation*}
where $\omega_{0}=\omega_{0}(T)\sim{T}$ are the zero-mode Matsubara frequencies of hot pQCD and $\epsilon_{v}<0$ is the vacuum energy density which can be, e.g. the bag constant of the MIT model.} in order to truly represent a dissipative term {in the relativistic hydrodynamic equations} \cite{Landau:1987b}. 
 
For NS matter{, i.e. cold and dense QCD matter}, some simple models studied this {in-medium} observable ${\Delta}$, e.g. Ref. \cite{Haensel:2007} found that {an expected} $\Delta \geq{0}$ is violated even while keeping Lorentz invariance because the short-range repulsion of the effective nuclear interactions tend to stiffen the EoS very rapidly at high densities. {Of course}, their conclusion {is of limited scope since it is valid only for one-phase stars, i.e. they do not consider the presence of phase transitions}. Nevertheless, in principle, strongly-discontinuous phase transitions might arise in dense QCD matter due to: i) generic arguments against the Schaefer-Wilczek conjecture (for a smooth hadron-quark continuity) posing the necessity of a discontinuous transition \cite{Cherman:2018jir,Cherman:2020hbe}, ii) potential effects from non-perturbative Gribov-Swanziger-like models for dense matter already favoring $\Delta < 0$ in SU(2) model at intermediate $T$ \cite{Canfora:2015yia}, iii) the activation-decoupling \cite{Jakobus:2020nxw,Grozin:2012ec} of hadronic and/or quark degrees of freedom (affecting $\alpha_{s}$) around the transition point, iv) the ``hidden'' pseudo-conformal QCD symmetry yet to be better understood when changing phases \cite{Rho:2022wco}, among other works \cite{Alvarez-Castillo:2018pve,Zacchi:2015oma,Zacchi:2016tjw}. 
  
In order to probe {the maximal effects of} this kind of {transitions on} dense matter, one might employ twin NSs \cite{Gerlach:1968zz,Schertler:2000xq,Alvarez-Castillo:2017qki} {as theoretical laboratories. Being more specific, these} unique stellar objects satisfy those strong-discontinuity requirements {due to the sharp interface (within the Maxwell construction at the transition point) separating the hadronic and quark phases \cite{Gerlach:1968zz,Schertler:2000xq,Alvarez-Castillo:2017qki}. It is worth to mention that these stars are known as ``twins'' (sometimes also called the third family/branch of compact stars) because for a given gravitational mass in the $M$-$R$ diagram, one could have two NSs with significantly different radii \cite{Gerlach:1968zz,Schertler:2000xq,Alvarez-Castillo:2017qki}. These radii differences are originated fundamentally in their internal chemical composition and thermodynamic construction \cite{Glendenning:2000}, being the less-compact (larger radius) star a purely hadronic one while the more-compact star (smaller radius) is called a twin.} 

Unfortunately, {in spite of being interesting by themselves}, there are currently few EoS studies (see, e.g. Refs. \cite{Brandes:2023hma,Brandes:2023bob,Annala:2023cwx} for interpolation, Bayesian and other agnostic approaches) dealing {with phase transitions in NSs. Most of these studies} conclude that it is very unlikely for this type of strong discontinuities to occur {in NSs and its corresponding EoS}. {Opposite to that,} other recent  works point out (for instance, Refs. \cite{Gorda:2022lsk,Christian:2023hez,Li:2024lmd} from NS hydrostatic observables and Refs. \cite{Blacker:2024tet,Hensh:2024onv} from fully dynamical NS mergers) that it is too early to dismiss this intriguing possibility which only an ab initio non-perturbative approach might conclusively discard {it} or not (see, e.g., Ref. \cite{Yamamoto:2021fjs} applying quantum computing to dense LQCD).

{On the other hand, the effects of a first-order phase transition on NS observables (adjusted to agree with up-to-date astrophysics constraints) were rigorously studied in} Refs. \cite{Takatsy:2023xzf,Albino:2024ymc} through the Bayesian approach, while Ref. \cite{Zhou:2024yzy} employed a deep-neural-network viewpoint. In particular, {all these works} indicate again that $\Delta$ is almost always positive, except within a small energy-density region which favors {a slightly negative and smooth behavior of $\Delta$} \cite{Takatsy:2023xzf} approaching pQCD {conformality} at high densities (see, e.g. Refs. \cite{Albino:2024ymc,Zhou:2024yzy} for the opposite behaviors). On this basis, no robust/clear explanation for this behavior {of $\Delta$ in hybrid stars is given and, more importantly, in twin stars. Unfortunately, some works (e.g., Ref. \cite{Albino:2024ymc}) are even currently employing the $\Delta>0$ conjecture \cite{Fujimoto:2022ohj} {as working principle} to reduce their parameter space. We believe that this is hasty if we lack a QCD proof reliably supporting that conjecture}. Moreover, although the recent work of Ref. \cite{Fujimoto:2024ymt} also attempts to probe the effects of a first-order transition on $\Delta$, the transition strength (jumps in $\Delta\epsilon$) depends upon the maximal Bayesian-inference thickness in `$\epsilon$' of their EoS band, which in practice is not large enough to produce twin stars with its novel implications for $\Delta$, as we do here.

Thus, it is necessary a comprehensive study of the behavior of $\Delta$ when strong discontinuities are present, which for us the perfect astrolaboratory will be twin stars. As far as we know, the only work studying a little bit about $\Delta$ in twin stars (in the context of modified theories of gravity) was that of Ref. \cite{Lope-Oter:2024egz}, although their findings pointed out to slightly negative values (minimally reaching $\Delta\approx -0.1$, a value which was also considered in the work of Ref. \cite{Fujimoto:2022ohj}) without proving the radial stability of their stars nor gravity-independent inference of $\Delta$. Apart from that work, only Refs. \cite{Ecker:2022dlg,Cai:2023pkt,Cai:2024oom} investigated $\Delta$ in some detail for a generic set of hybrid NS EoSs with continuous transitions finding $\Delta \gtrsim 0$ connected to their maximal $c^{2}_{s}$ ($\sim 0.8$). Notice that their assumption of slightly negative $\Delta$ was reached monotonically (without restriction due to a possible change of phases) approaching very slowly $\Delta\to 0$ to agree with pQCD {at high $\mu_{B}$}.

In the present work, we shall focus our attention on the behavior of the normalized trace anomaly, $\Delta$, in cold/dense matter suffering a strong discontinuous phase transition producing twin NSs in a large parameter space. The paper is organized as follows. In Secs. \ref{sec:twinmatter}, \ref{sec:Tanomaly} we explain the main ideas behind twin stars, i.e. hydrostatic and dynamical conditions for their existence, EoSs and behavior of $\Delta$ with discontinuous transitions. Section \ref{sec:TSanom_I} presents our robust findings for $\Delta$ within the distinct categories of studied twin stars. Rapid and slow conversions are also considered when characterizing the sign of $\Delta$. In Sec. \ref{sec:disc} we give analytical discussions for our numerical findings in Sec. \ref{sec:TSanom_I} to get insights for the appearance of negative $\Delta$ around the transition point. Section \ref{sec:conclusion} presents our conclusions and outlook. Additional discussions are presented in the five Appendices. {Throughout this work, we use natural units $G=c=1$ when displaying our methodology (e.g. $\epsilon=\rho$ in equations or EoSs, with `$\rho$' as the rest mass density) but returning their physical units for the analysis in the main text and plots.}

\section{Twin-star essentials}
\label{sec:twinmatter}

{In} this section, we {will} point out the {main physical conditions concerning the stability of relativistic stars with 1st-order transitions \cite{Pereira:2017rmp}, i.e.}  the stringent Seidov's criterium and radial stability of the fundamental mode with rapid/slow junction conditions, respectively, being now applied to twin neutron stars. In practice, these conditions will serve us to properly choose the parameters in our EoSs for the hadronic and quark phases {in order to get fully stable twin stars.} {After that, we will be able to probe} the behavior of $\Delta$ {in a robust manner}.

\subsection{SEIDOV CRITERIUM\\ AND TWIN-STAR MATTER}

{Regarding the microphysics input, the hybrid EoS giving rise to} twin NSs families {can be built} from a matching between a hadronic EoS (e.g. coming from CET satisfying low-density nuclear data) up to a transitional pressure, $P_{t}$, from which a QM EoS begins through a strongly discontinuous phase transition, both phases connected by a Maxwell or Gibbs construction \cite{Gerlach:1968zz,Schertler:2000xq,Alvarez-Castillo:2017qki,Glendenning:2000}. {In particular, considering the former case} at the transition pressure, the energy and baryon number densities present large jumps ($\Delta \epsilon$ and $\Delta n_{B}$, respectively) manifested through a finite value of latent heat $Q^{*}\equiv \epsilon^{\rm min}_{Q}-\epsilon^{\rm max}_{H}=\mu_{c}\Delta n_{B}$ (being `$\mu_{c}$' the critical baryochemical potential, `$\epsilon^{\rm min}_{Q}$' the minimum value of energy density for the QM phase {after the transition}, and `$\epsilon^{\rm max}_{H}$' the maximum value reached by the hadronic phase {before the transition}). Besides, it should always be kept in mind that only a delicate matching of a stiff hadronic EoSs to a stiff QM one ($c^{2}_{s}>1/3$) favors the existence of twin hybrid NSs \cite{Alvarez-Castillo:2017qki,Jakobus:2020nxw,Alvarez-Castillo:2018pve}.

\begin{figure*}[!t]
	\vspace*{-0.55cm}
  \hspace*{-0.65cm}
  \includegraphics[width=0.54\textwidth]{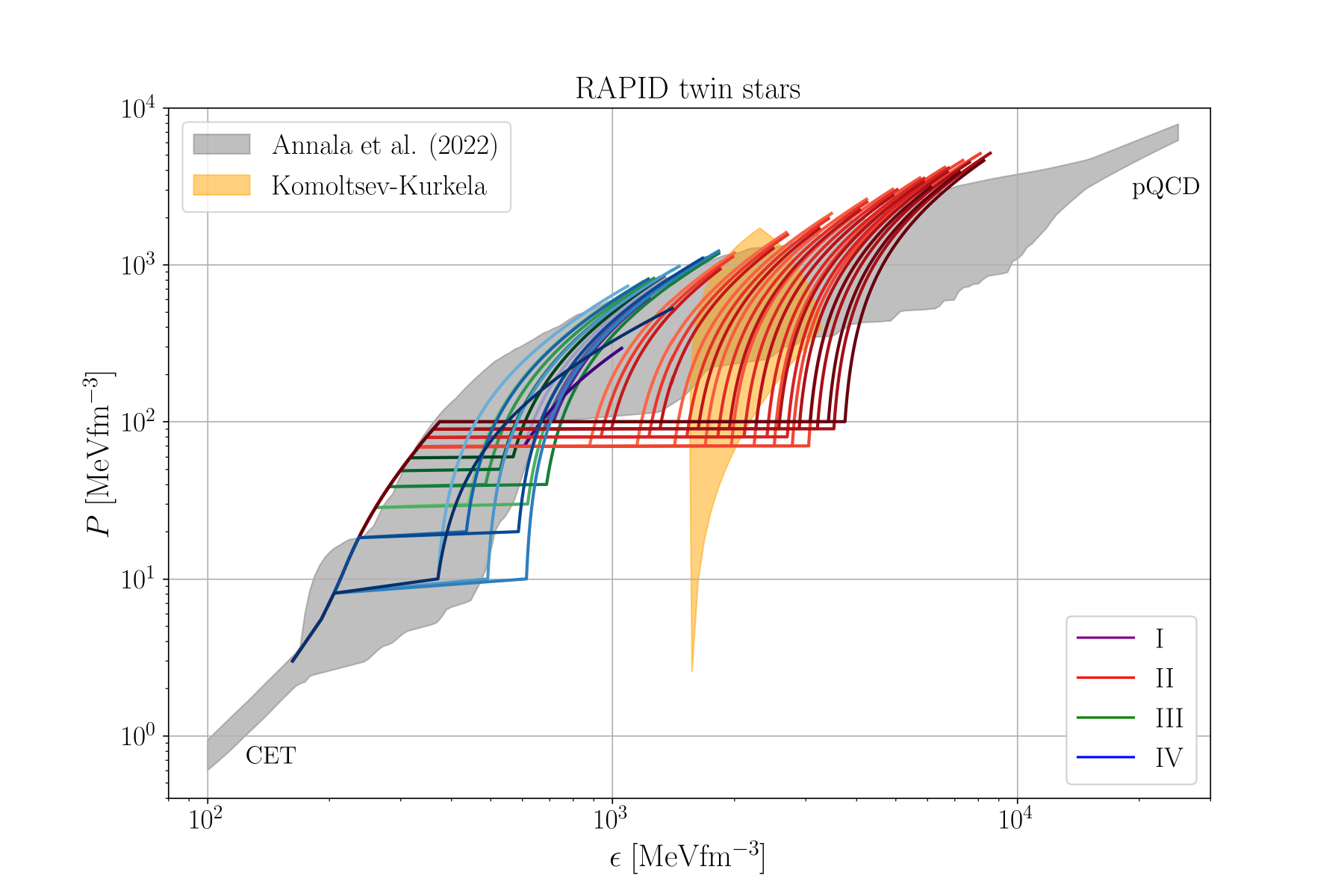}
  \hspace*{-1.1cm}
  \includegraphics[width=0.54\textwidth]{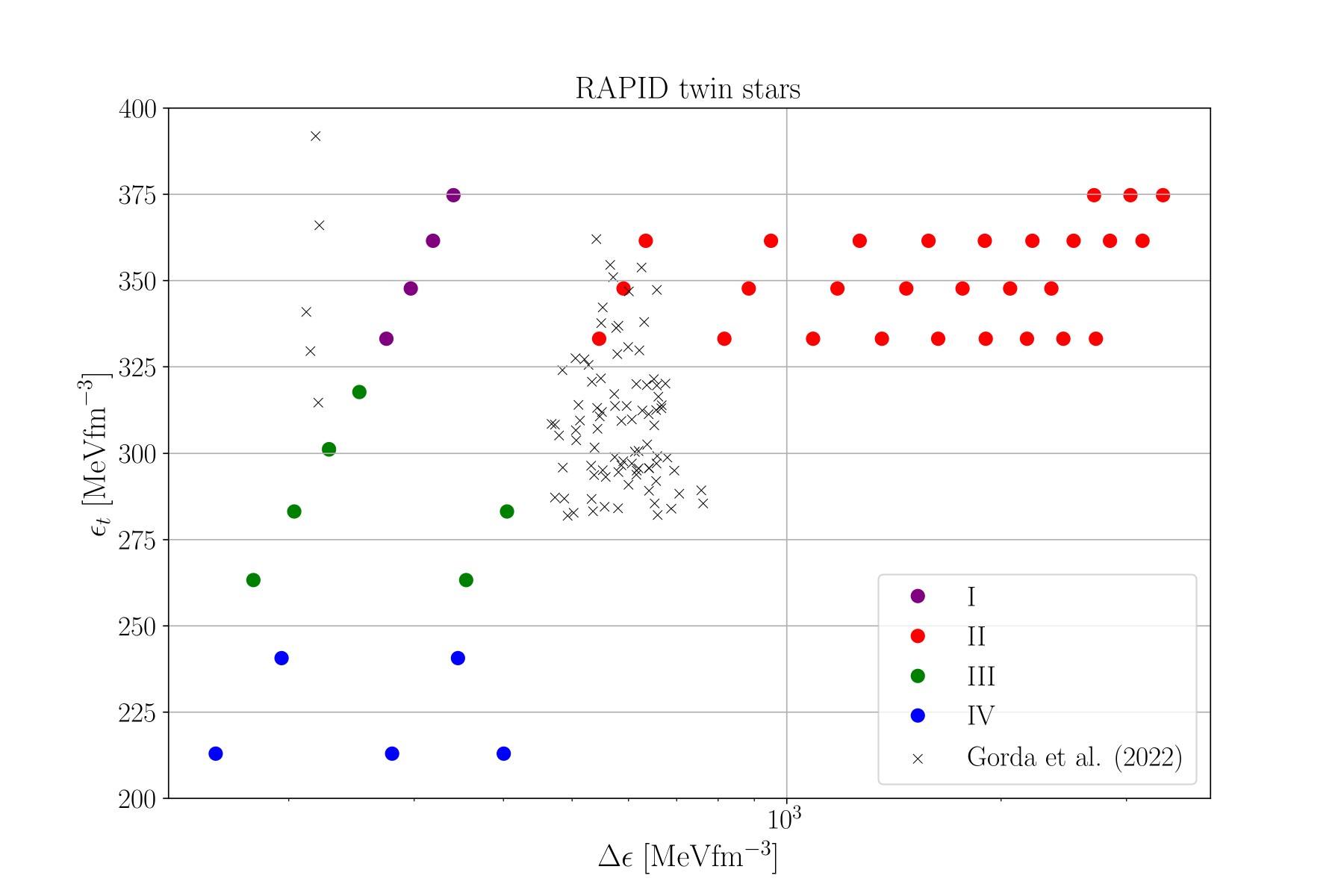}
  \caption{\label{fig:stiff_rapid_eos} Family of {twin} EoSs ({\it left panel}) and related Seidov's parameter space ({\it right panel}) for Category I-IV stable twin stars with rapid conversions. For comparison, we also display ({\it left panel}) results from CET~\cite{Hebeler:2013nza} and pQCD~\cite{Kurkela:2009gj} as well as the astrophysics-constrained band (grey) of EoSs of Annala et al.~\cite{Annala:2021gom} at intermediate densities. {We also add the} pQCD constraint (orange band) of Komoltsev-Kurkela~\cite{Komoltsev:2021jzg} {at $n_{B}=10n_{0}$}. Also, the symbols `$\times$' ({\it right panel}) represent the findings of Ref. \cite{Gorda:2022lsk}.}
\end{figure*}

{\bf Seidov criterium:} {As announced above,} for twin stars to exist, they must satisfy the Seidov criterion \cite{Alvarez-Castillo:2017qki}, which establishes the
critical value in the energy density jump, $\Delta\epsilon_{\rm crit}$, in the sense that for smaller values, the hybrid {(twin)} and hadronic branches are connected, and no twin stars are possible. This criterion is given by \cite{Alvarez-Castillo:2017qki}
\begin{equation}
  \label{eq:Seidov}
  \Delta\epsilon \geq \Delta \epsilon_{\rm crit} \equiv \frac{1}{2}\epsilon_t + \frac{3}{2}P_t,
\end{equation}
being `$\epsilon_{t}$' the onset energy density for the phase transition to occur. {In fact, in our case of twin stars, $\epsilon_t$ corresponds to the last and maximum (since $P$ is a monotonically increasing function $\epsilon$) value of the hadronic phase, i.e. $\epsilon_t=\epsilon^{\rm max}_{H}$ (one can see this in the left panel of Fig.~\ref{fig:stiff_rapid_eos} and some of its taken values in Table \ref{tab:css_parameters}).} Thus, Eq. (\ref{eq:Seidov}) ensures the presence of an unstable branch in the $M$--$R$ diagram, giving rise later to the ultradense and stable branch {known as} {\it twin stars}. We now model their EoSs.

{\bf Parameters at the transition point:} Our twin EoSs will have $P_{t}$'s in the range [10, 100]~MeVfm$^{-3}$ for which we computed $\Delta\epsilon$ varying it between 1.1 and 10 times its initial value. Since twin stars require a stiff QM EoS, we probe two high values for its respective $c^{2}_{s}$, the first $c^{2}_{s}=1$ and the second $c^{2}_{s}=0.5$. Notice that both are non-conformal, {the former} reaching the causality limit.

{\bf Hadronic phase:} {This confined phase will be} conveniently described (from a numerical viewpoint) by the generalized piecewise polytropes (GPP)~\cite{OBoyle:2020qvf} that consistently assures the continuity of the pressure, energy density and their derivatives. For the outer crust, we use the GPP form of the SLy4 (as presented in Table III of Ref.~\cite{OBoyle:2020qvf}). For densities larger than 1.1$n_0$ (being $n_0=0.16~{\rm fm^{-3}}$ the saturation density), we will assume a model agnostic GPP that connects with the stiffest CET EoS~\cite{Lugones:2021bkm,Goncalves:2022ymr}. In order to achieve this, our model agnostic GPP has the following parameters\footnote{{For more details on these parameters and the next segments of the agnostic GPP, we refer the reader to Sec. IV of Ref. \cite{OBoyle:2020qvf}.}}:
$\log_{10}[\rho_0 /({\rm g~cm^{-3}})] = 13.865$, $\log_{10} K_1 = -27.22$, $\Gamma_1 = 2.77$, $\Gamma_2 = 7.0$ and $\Gamma_3 = 3.0${, where `$\rho_0$' represents the starting mass density, $K_1$ is the first polytropic constant and $\Gamma_i$ are the exponents of the rest mass density `$\rho$'}. For the other two dividing densities, we have used $\log_{10}[\rho_1 /({\rm g~cm^{-3}})] = 14.45$ and $\log_{10}[\rho_2 / ({\rm g~cm^{-3}})] = 14.58$. It is worth to stress at this point that in Appendix B we display our findings for $\Delta$ employing the so-called {\it intermediate} CET of Ref. \cite{Hebeler:2013nza} in order to probe the robustness of our conclusions. 

{\bf Quark phase:} For this high-density sector of the twin EoS, we employ the well known constant-speed-of-sound (CSS) parametrization \cite{Alford:2013aca}, which allows the description of the hybrid EoS through a set of three parameters: $P_{t}$, $\Delta \epsilon$ and constant {speed of sound for the quark phase,} $c^{2}_{s, Q}\equiv s$. In this manner, the full hybrid EoS is written as (with Maxwell construction)
\begin{equation}
  \label{eq:css_parametrization}
  \epsilon(P) = \begin{cases}
    \epsilon_{\mathrm{H}}(P) \hfill P < P_{t} \,, \\
    \epsilon_{\mathrm{H}}(P_t) + \Delta\epsilon + s^{-1}(P - P_{t}) \hspace{.5cm} P > P_{t}\,.
  \end{cases}
\end{equation}
For illustrative purposes, we present in Table \ref{tab:css_parameters} some particular values for these needed parameters to construct a CSS EoS for twin stars and estimate their magnitudes. For this, we follow the classification of Ref. \cite{Christian:2017jni} (see also references therein) of four types of twins depending basically on the combination of maximal masses obtained for the hadronic and twin branches within each category, i.e. Category I: the hadronic and twin maxima exceed the $2M_{\odot}$ limit, Category II: only the hadronic maximum reaches the $2M_{\odot}$ limit, Category III: the maximum of the hadronic branch lies in the range $[1, 2]M_{\odot}$ while the twin maximum exceeds $2M_{\odot}$, and Category IV: the twin maximum exceeds the $2M_{\odot}$ limit but the maximum of the hadronic branch appears below $1M_{\odot}$.

In the left panel of Fig.~\ref{fig:stiff_rapid_eos}, we present our employed twin EoSs. {Please note that each category is represented by a distinct color, which will be consistent throughout this work.} Being more specific, in the next subsection, we will classify two types of twins depending on their stability against radial pulsations: the rapid and slow conversions. In turn, this allows us to construct hundreds of twin EoSs which, unfortunately, might obscure the visual analysis of their behavior. Thus, without loss of generality, we chose to display only the results for the rapid case which as a whole is a somewhat smaller set compared to the slow twin stars whose EoSs overlap several times and imposes visual challenges for the analysis. {This is crucial when comparing our EoSs} with the broad grey band of EoSs (only adjusted to produce NSs with masses higher than $2M_{\odot}$ and the GW170817 tidal deformability limit) of Annala et al. \cite{Annala:2021gom} (containing also results for the extreme opposite limits of CET and pQCD) which for us represents a benchmark NS EoS, since they explicitly did not considered the possibility of a first-order transition for their intermediate-density interpolations. 

\begin{table}
  \begin{tabular}{c|c|c|c|c|c}
    \hline
    Category & $\epsilon_{H}^{\rm max}=\epsilon_{t}$ & $\epsilon_{Q}^{\rm min}$ & $P_{t}$ & $\Delta\epsilon$ & $c^{2}_{s}$  \\
    \hline
    I & 333.08 & 607.34 & 70 & 274 & 1 \\
    \hline
    II & 333.08 & 878.88 & 70 & 545 & 1 \\
    \hline
    III & 263.73 & 441.62 & 30 & 178 & 1 \\
    \hline
    IV & 212.91 & 370.85 & 10 & 157 & 1
  \end{tabular}
  \caption{\label{tab:css_parameters}Sample of parameters used in this work for the CSS parametrization. All they are in MeV\,fm$^{-3}$ while the speed of sound is in units of the speed of light.}
\end{table}

Besides, in this same {left panel of Fig. \ref{fig:stiff_rapid_eos},} we put the orange band of Komoltsev-Kurkela \cite{Komoltsev:2021jzg}. {This} is a region of pressures and energy densities {at around $10n_{0}$,} which can be consistently (from the thermodynamic viewpoint) and very generically ({through} any reasonable smooth curve) {ensured to approach} pQCD at $40n_{0}$. In practice, if our twin EoS pass through the orange band, we have some certainty of expecting that our EoSs will reach pQCD at $40n_0$. On the other hand, although the study of Komoltsev-Kurkela considered also potential first-order transitions, they maximally focused on the behavior of the NS EoS at $10n_0$. Nevertheless, one can clearly see in the left panel of Fig. \ref{fig:stiff_rapid_eos} that Categories I and II lie at even higher densities, {i.e. around $10\leq n_{B}/n_{0} \lesssim 30$}, thus leading us to believe that the work of Komoltsev-Kurkela should be extended to the twin-star extreme densities at that density region. Interestingly, Categories III and IV fit very well to the band of Annala et al. \cite{Annala:2021gom}. {On the other hand}, in the right panel of Fig. \ref{fig:stiff_rapid_eos}, we present a comparison {between} our parameter space (satisfying Eq. \ref{eq:Seidov}) {and} the one from Ref. \cite{Gorda:2022lsk}, where we show the energy density at the transition point $\epsilon_t$ as a function of $\Delta\epsilon$. {Notice that} our values {for the energy density jumps $\Delta\epsilon$} of Category II are significantly larger than those in Ref.~\cite{Gorda:2022lsk}.

\subsection{STABLE TWIN STARS: \\RAPID AND SLOW CONVERSIONS}

{\bf Hydrostatic equilibrium:} {The structural properties of spherical and static} NSs{, e.g. its $M$-$R$ relations, are obtained after solving} the TOV equations {for the hydrostatic equilibrium between the fluid's expansion and compressive gravitational pull \cite{Glendenning:2000}.} They are given by:
\begin{equation}
\frac{dP}{dr}=-\frac{\epsilon\mathcal{M}}{r^{2}}\left(1+\frac{P}{\epsilon}\right)\left(1+\frac{4\pi{r^3}{P}}{\mathcal{M}}\right)
  \left(1-\frac{2\mathcal{M}}{r}\right)^{-1} \; ,
\label{TOV1}
\end{equation}
\begin{equation}
\frac{d\mathcal{M}}{dr}=4\pi{r}^{2}\epsilon,~~\frac{d\nu}{dr}=-\frac{2}{P+\epsilon}\frac{dP}{dr} \; ,
\label{TOV2}
\end{equation}
where $P$ is the pressure, $\epsilon$ is the energy density, $\mathcal{M}$ the gravitational mass at radius $r$, and $\nu(r)$ the temporal metric function in the Schwarzschild-like metric. {To solve these equations, one imposes conditions for the pressure and mass variables at the stellar center until a boundary is found when $P(r=R)=0$, thus allowing us to get a value of gravitational mass $\mathcal{M}(r=R)\equiv M$ and stellar radius $R$ \cite{Glendenning:2000}. This serves us to draw the $M$-$R$ relations.}

{\bf Dynamical equilibrium:} {After having found the compact-star hydrostatic configurations from the TOV equations, one could ask if they are stable against (periodic and infinitesimal) radial\footnote{{Interestingly, it has been proven \cite{Passamonti:2005cz,Passamonti:2007tm} that radial modes are related to non-radial ones non trivially, thus affecting the gravitational-wave spectrum coming from isolated or binary NSs.}} oscillations \cite{Glendenning:2000}. This is an stringent requirement to test if these stars are really realized in nature as realistic configurations or not. Historically, this problem was first studied by S. Chandrasekhar producing its famous Sturm-Liouville problem \cite{Glendenning:2000} for the radial perturbations. After solving these equations, it is usual to classify a set NS configurations as {\it dynamically stable} if they have a positive eigenfrequency of the zero mode, i.e $\omega^{2}_{n=0}\geq 0$. This classification is consistent even if we are not dealing with explicit time dependencies \cite{Ruoff:2000nj}, i.e. perturbative expansions were applied around the hydrostatic TOV solutions in the fully dynamical general-relativistic equations, then yielding exactly the Chandrasekhar's eigenvalue problem.}

{Besides, the aforementioned classification helps us to clearly distinguish it from the widely spread and simple {\it static} criterium of stability: $dM/d\epsilon_{c}\geq 0$ (with $\epsilon_{c}$ the central energy density), which must be applied directly to the TOV solutions \cite{Glendenning:2000}. In few words, this criterium states that as long as one has a positive derivative, we have stable stars. In particular, the extreme configuration satisfying $dM/d\epsilon_{c}\to 0$ represents the maximally stable star. In this sense, a maximally-stable star (for a given EoS) could be understood as a transition configuration between stable and unstable stars. Interestingly, this criterium has been proven to be equivalent to the criterium from the radial-oscillation viewpoint for one-phase stars \cite{Glendenning:2000}, i.e. as one approaches $\omega^{2}_{n=0}\to 0$, the maximally-stable is reached. However, this equivalence changes non trivially for twin NSs, since auxiliary conditions should be imposed in the radial pulsation equations at the hadron-quark interface.}

{Before particularizing our framework to twin NSs, we establish our employed set of radial pulsation equations. As usual in the recent literature, we will solve the Gondek's equations for physical clarity and numerical convenience.} They are a pair of coupled first-order differential equations \cite{Gondek:1997fd} for the relative radial displacement $\Delta{r}/r\equiv\xi$ and Lagrangian perturbed pressure $\Delta{P}$ (implicitly having factors of ${e}^{i\omega_n{t}}$). This problem can be compactly written in matrix form as \cite{Gondek:1997fd}:
	\begin{equation}
  \begin{pmatrix}
   \dfrac{\strut d\Delta{P}}{\strut dr} \vspace{0.2cm}
    \\
    \dfrac{\strut d\xi}{\strut dr}
  \end{pmatrix}
  =
  \begin{pmatrix}
    \hspace{0.1cm}\mathcal{Z}(r) & \hspace{0.2cm} \mathcal{Q}(r, \omega^{2}_{n}) \vspace{0.9cm}
    \\
    \hspace{0.1cm}\mathcal{R}(r)\vspace{0.2cm} & \hspace{0.2cm} \mathcal{S}(r)
  \end{pmatrix}
  \begin{pmatrix}
    \Delta{P} \vspace{0.9cm}
    \\
    \vspace{0.3cm}\xi
  \end{pmatrix} \;,
\end{equation}
where $\omega^2_{n}$ is the oscillation eigenvalue and $\mathcal{Z}$, $\mathcal{Q}$, $\mathcal{R}$, $\mathcal{S}$ are radial functions giving by \cite{Gondek:1997fd}
\begin{equation}
\mathcal{Z}(r)=-4{\pi}r{e}^{\lambda}(\epsilon+P)+\frac{1}{\epsilon+P}\frac{dP}{dr},
\end{equation}
\begin{multline}
\mathcal{Q}(r,\omega^{2})=-8{\pi}rP(\epsilon+P){e^{\lambda}}+ \\ \frac{r}{\epsilon+P}\left(\frac{dP}{dr}\right)^{2}-4\frac{dP}{dr}+{\omega^{2}r(\epsilon+P)e^{\lambda-\nu}},
\end{multline}
\begin{equation}
\mathcal{R}(r)=-\frac{1}{r}\frac{1}{P\Gamma},\hspace{0.5cm} \mathcal{S}(r)=-\frac{1}{\epsilon+P}\frac{dP}{dr}-\frac{3}{r},
  \label{Eq.Gondek}
\end{equation}
being $\Gamma(r)=(1+\epsilon/P)c^{2}_{s}$ the adiabatic-index profile and $\lambda(r)=-\ln(1-2\mathcal{M}/{r})$ the spatial metric function of the Schwarzschild-like metric inside the neutron star. {Basically, the numerical convenience of this framework comes from the absence of radial derivatives of $\Gamma$ \cite{Gondek:1997fd}.}

{\bf Rapid and slow junction conditions:} {As already announced, a careful discussion on the dynamical equilibrium of twin stars is in order. The main difference with one-phase stars comes from the fact that these ultradense stars have a hadronic mantle with a somewhat large QM core. In other words, first-order phase-transition discontinuities in the EoS require immediately the addition of auxiliary boundary conditions at the interface. They are usually dubbed {\it junction conditions} (JC). From the microscopic perspective, they arise because the potential exchange of matter at the hadron-quark phase-splitting interface of coexistence within these twin stars due to strong and weak nuclear chemical reactions \cite{Haensel:1989wax,Pereira:2017rmp}. Besides, depending on the timescale of these reactions, we classify the extreme cases as {\it rapid (slow)} if they are very small (large) compared to the radial-oscillation period of the fundamental mode ($n=0$) \cite{Haensel:1989wax,Pereira:2017rmp}. In Table \ref{tab:BCs1} we list these JC for the slow and rapid conversions at the phase-transition point of the QCD deconfinement transition here modeled as being strongly discontinuous.}

\begin{table}[t]
  \begin{center}
    \begin{tabular}{c|c|c} 
${\rm JC}$ & Rapid & Slow\\
      \hline
$\xi(r=r_{\rm inter})$  &$\left[\xi - {\Delta P}/({r P'})\right]^{+}_{-}=0$ &$[\xi]^{+}_{-}=0$ \\
$\Delta P(r=r_{\rm inter})$ & $[\Delta P]^{+}_{-}=0$&$[\Delta P]^{+}_{-}=0$\\
        \end{tabular}
        \caption{Radial oscillation junction conditions (JC) at the interface ($r_{\rm inter}$) for rapid and slow conversions. Note that $[\xi]^{+}_{-}\equiv \xi^{+}-\xi^{-}$ and $[\Delta P]^{+}_{-}\equiv \Delta P^{+}-\Delta P^{-}$, meaning +(-) to the right (left) of the interface with increasing $r$. Besides, $P'\equiv dP/dr$, where `$P(r)$' is a solution of the TOV equations.}
      \label{tab:BCs1}
  \end{center}
\end{table}

{Now,} in order to {simultaneously solve the} system of coupled equations, Eqs. (\ref{TOV1})--(\ref{Eq.Gondek}), {determining the radial eigenfrequencies, we first solve the TOV equations and use its solutions (providing the thermodynamic and metric-function profiles) as inputs in the Gondek's equations. We list the required boundary conditions in Table \ref{tab:BCs2}, where the main input is the twin NS EoS, but now appropriately written as $\epsilon=\epsilon(P)$ for numerical purposes since it contains a first-order transition separating the nuclear and QM phases with a finite jump $\Delta \epsilon$ at the transition point. This choice spoils our numerical code if `$\epsilon$' is the independent variable. Instead, `$P$' is continuous at all densities even with the strong transition.}

\begin{table}[t]
  \begin{center}
    \begin{tabular}{c|c|c|c|c|c|c} 
${\rm BC}$ & $P$ & $\mathcal{M}$ & $\nu$ & $\lambda$ & $\xi$ & $\Delta P$\\
      \hline
$r=0$  &$P_{0}$ &$0$ &$\nu_{0}$ & 0 & 1 & $-3(\xi{P}\Gamma)_{\rm center}$\\
$r=R$ & $0$&$M$ &$\ln\left(1-{2M}/{R}\right)$ & -$\nu(R)$ & {\rm finite} & 0\\
        \end{tabular}
        \caption{Set of boundary conditions (BC) at the NS center ($r=0$) and surface ($r=R$) to solve simultaneously the TOV+Gondek's equations, with `$\nu_{0}$' chosen appropriately.}
      \label{tab:BCs2}
  \end{center}
\end{table}

{All this procedure will give us $\omega^{2}_{n=0}\equiv \omega^{2}_{n=0}(P_{0}, M)$, i.e. the fundamental-mode eigenfrequencies
for a star with central pressure $P_{0}$ and gravitational mass $M$. As long as $\omega^{2}_{n=0}>0$, the corresponding twin star will be said to be dynamically stable for the rapid and slow JC. In particular, the configuration satisfying $\omega^{2}_{n=0}\to 0$ will determine the maximal dynamically-stable star with corresponding maximal central pressure, $P^{\rm max}_{0}$, and mass $M_{\rm max}$ for a given EoS. This compact star marks the end of stable configurations, then initiating the unstable sector of the whole sequence characterized by having $\omega^{2}_{n=0} <0$ implying exponentially increasing oscillation amplitudes leading to violent explosions or gravitational collapse to a black hole.}

{At this point, it is worth to mention that the static stability criterium ($dM/d\epsilon_{c} \geq 0$) agrees with results for the radial oscillation analysis in twin stars only for the rapid JC. For the case of slow JC, one can have twin stars with $dM/d\epsilon_{c} < 0$ but still $\omega^{2}_{n=0} > 0$. Since the static criterium is derived from the radial-pulsation solutions, we choose solutions of the last method as conclusive, then leaving aside the static criterium.}

{\bf Reaction modes in twin stars?} {An interesting phenomenon occurs} in the case of rapid JC, one extra highly non-trivial radial mode appears, the so-called {\it radial reaction mode}, which is present only when strong transitions occur at NS interiors \cite{Pereira:2017rmp}. This mode becomes relevant when {it is} the fundamental one with eigenfrequency $\omega^{2}_{n=0}\equiv \omega^{2}_{R}$. {It should be noticed that only} delicate combinations of $P_{t}$, $\Delta \epsilon$ and $\eta\equiv \epsilon^{\rm min}_{\rm Q}/\epsilon^{\rm max}_{\rm H}$, i.e. $\omega^{2}_{R}\sim (3[1+P_{t}/\epsilon^{\rm max}_{\rm H}]-2\eta)/(\eta-1)$ \cite{Pereira:2017rmp} {favors its appearance. In sections below, we will probe its presence in our twin stars for the four categories explored.}

\section{The QCD trace anomaly \\with strong transitions}
\label{sec:Tanomaly}
{\bf Overview:} A few years after asymptotic freedom was discovered within pQCD, it was realized non-perturbatively in Refs. \cite{Collins:1976yq,Nielsen:1977sy} that (even by assuming massless quark flavors), the trace of the energy-momentum tensor of QCD, i.e. $\eta_{\mu\nu}T^{\mu\nu}_{\rm QCD}$ (being `$\eta_{\mu\nu}$' the Minkowski metric tensor) posses a so-called {\it quantum anomaly} associated to the dynamical breaking of scale-conformal invariance induced purely by the highly non-trivial nature of the real QCD vacuum. If one turns on the quark masses, the aforementioned trace passes to include an explicit breaking directly connected to the quark condensates. These results are \cite{Collins:1976yq,Nielsen:1977sy}
\begin{equation}
\label{eq:trace}
\eta_{\mu\nu}T^{\mu\nu}_{\rm QCD}\equiv T^{\mu}_{\mu}=\frac{\beta_{\rm QCD}}{2g}G^{a}_{\mu\nu}G^{\mu\nu}_{a}+(1+\gamma_{m})\sum_{f}m_{f}\overline{q}_{f}q_{f},
\end{equation}
where $\left\lbrace G^{a}_{\mu,\nu},q_{f} \right\rbrace$ are the renormalized gluon and quark fields, $\left\lbrace m_{f},g \right\rbrace$ the renormalized quark masses and gauge coupling, `$\beta_{\rm QCD}$' is the QCD beta function and `$\gamma_{m}$' is the anomalous dimensions of the quark mass. {It should be noted} that Eq. (\ref{eq:trace}) is valid for perturbative as well as non-perturbative energy scales. Besides, the behavior of this anomaly depends upon the number of active flavors. It is interesting to note that as of today, only pQCD results for `$\beta$' and `$\gamma_{m}$' are reliably known \cite{Ioffe:2010}.

{\bf In-medium case:} If one is interested in the equilibrium thermodynamics of Eq. (\ref{eq:trace}), it is appropriate to separate the vacuum and in-medium (dense and thermal) contributions in order to build the ensemble average {(denoted by $\left\langle ... \right\rangle$ at a given in-medium condition)} of the non-vacuum sector as
\begin{equation}
  \left\langle T^{\mu}_{\mu} \right\rangle_{\mu_{B},T}=\epsilon-3P.
\end{equation}
According to Ref. \cite{Fujimoto:2022ohj}, one can work more appropriately with its normalized version given by\footnote{With this definition, one can see explicitly its scale-invariant nature, i.e. defining the dimensionless $P'=P/\epsilon^{*}$ and $\epsilon'=\epsilon/\epsilon^{*}$, one gets an adimensional quantity $\Delta=1/3 -(P{'}\epsilon^{*})/(\epsilon{'} \epsilon^{*}) = \Delta'= 1/3 - P{'}/\epsilon{'}$, being $\epsilon^{*}$ some appropriate energy-density scale.}
\begin{equation}
  \label{eq:trace_anomaly}
  \Delta \equiv \frac{\left\langle T^{\mu}_{\mu} \right\rangle_{\mu_{B},T}}{3\epsilon} =  \frac{1}{3} - \frac{P}{\epsilon}\,.
\end{equation}
Interestingly, it was proven in Ref. \cite{Fujimoto:2022ohj} that this quantity has the extreme values between ($T=0$ and $\mu_{B} \neq 0$)
\begin{equation}
-\frac{2}{3}(\approx -0.667) \leq \Delta < \frac{1}{3}(\approx 0.333),
\end{equation}
the upper bound coming from the non-relativistic limit ($P\ll \epsilon$) and the lower bound from the causality bound ($P\leq \epsilon$). The conformal (ultra-relativistic) limit lies at $\Delta=0$ {when $P=(1/3)\epsilon$}. 

{Interestingly, Ref. \cite{Fujimoto:2022ohj} stated that one can relate $\Delta$ to} $c^{2}_{s}$ {by separating this last one in two terms as follows:}
\begin{equation}
c^{2}_{s}=c^{2}_{\rm s,~deriv}+c^{2}_{\rm s,~nonderiv},
\end{equation}
where these new terms (derivative and non-derivative) are defined as \cite{Fujimoto:2022ohj}:
\begin{equation}
\label{eq:speedT}
c^{2}_{\rm s,~deriv}\equiv -\epsilon \frac{d\Delta}{d\epsilon},\hspace{0.5cm} c^{2}_{\rm s,~nonderiv}\equiv\frac{1}{3} - \Delta,
\end{equation}
being the conformality of the system reached when $\Delta \to 0$ and $d\Delta/d\epsilon \to 0$, i.e. $c^{2}_{s}\simeq c^{2}_{\rm s,~nonderiv} \to 1/3$.

{\bf Behavior of `$\Delta$' at the quark-hadron interface:} Now, for our investigation of $\Delta$ in the interior of NSs with first-order phase transitions, we {start by solving} Eq. (\ref{eq:speedT}) with our hadronic and QM EoSs after finding $c^{2}_{s}$ for each phase. {For instance, it is} numerically straightforward {to compute $\Delta_{\rm H (Q)}$ from $c^{2}_{s, H (Q)}=dP_{H (Q)}/d\epsilon_{H (Q)}$ for the hadronic (quark) phase knowing a priori the EoSs.}

{On the other hand, the quark-hadron interface (QHI)} satisfies $c^{2}_{s}=0$ (since $P_{t}$ is constant within the Maxwell construction), {then forcing} Eq. (\ref{eq:speedT}) to satisfy the next differential equation with respective initial condition, i.e. the normalized trace anomaly evaluated at the maximal hadronic density before the transition, $\Delta^{\rm max}_{H}$ (see Sec. \ref{sec:twinmatter} for discussion):
\begin{equation}
\epsilon\frac{d\Delta_{\rm QHI}}{d\epsilon}+\Delta_{\rm QHI}=\frac{1}{3},\hspace{0.5cm} \Delta_{\rm QHI}(\epsilon^{\rm max}_{H})\equiv \Delta^{\rm max}_{H},
\end{equation}
with general solution given by
\begin{equation}
\label{eq:deltaMix}
\Delta_{\rm QHI}(\epsilon)=\frac{1}{3}\left(1-\frac{\epsilon^{\rm max}_{H}}{\epsilon}\right)+\frac{\epsilon^{\rm max}_{H}}{\epsilon}\Delta^{\rm max}_{H},
\end{equation}
which in turn leads us to obtain $c^{2}_{\rm s,~deriv,QHI}$ as

\begin{equation}
\label{eq:csMix}
c^{2}_{s\rm ,~deriv,QHI}=\frac{\epsilon^{\rm max}_{H}}{\epsilon}\left(\Delta^{\rm max}_{H}-\frac{1}{3}\right)=-c^{2}_{s\rm ,~nonderiv,QHI}.
\end{equation}

{Since} one should be careful with the interpretation of Eqs. (\ref{eq:deltaMix})--(\ref{eq:csMix}), {we pass to discuss them in some detail}. Firstly, $\Delta_{\rm QHI} (\epsilon)$ cannot be considered physical within our Maxwell construction and it should be understood as an artifact that joins $\Delta_{H}$ to $\Delta_{Q}$ in a $\Delta$ vs $\epsilon$ plane. Besides, it will be clear from our figures below that in all cases $\Delta_{\rm QHI}$ is an smooth increasing curve (strictly speaking, a branch of a negative rectangular hyperboles) although we do not display their explicit behavior keeping that region in blank. In few words, the $\Delta \epsilon$ jump induces a smooth jump in $\Delta(\epsilon)$. Interestingly, this behavior was also found in Ref. \cite{Zhou:2024yzy} within their deep-neural network calculations but without {explicit} physical explanation. Secondly, {from} Eq. (\ref{eq:csMix}) {one can see} the opposite nature of these {derivative and non-derivative} terms at the quark-hadron interface which for us is unphysical. {However, both speed-of-sound terms will have a different and independent behavior within the Gibbs/Glendenning construction~\cite{Glendenning:2000}. This happens because in the mixed phase $c^2_s \neq 0$, i.e. it is density dependent. We leave this case for future study.}

{\bf Onset to the QM phase:} After the strong transition occurs, the $\Delta_{Q}$ for the QM phase (with $c^{2}_{s, Q}= \text{const} \equiv s$ within our CSS parametrization) from Eq. (\ref{eq:speedT}) produces
\begin{equation}
\epsilon\frac{d\Delta_{Q}}{d\epsilon}+\Delta_{Q}=\frac{1}{3}-s,\hspace{0.5cm} \Delta_{Q}(\epsilon^{\rm min}_{Q})\equiv \Delta^{\rm min}_{Q},
\end{equation}
the last term being the initial boundary condition at the minimal value of QM energy density after the transition {occurs, $\Delta^{\rm min}_{Q}$, then} having the following general solution
\begin{equation}
\label{eq:deltaQM}
\Delta_{Q}(\epsilon)=\left(\frac{1}{3}-s\right)\left(1- \frac{\epsilon^{\rm min}_{Q}}{\epsilon}\right)+\frac{\epsilon^{\rm min}_{Q}}{\epsilon}\Delta^{\rm min}_{Q},
\end{equation}
which {in turn} leads to
\begin{equation}
\label{eq:csQM}
c^{2}_{s{\rm,~deriv},~Q}=\frac{\epsilon^{\rm min}_{Q}}{\epsilon}\left(\Delta^{\rm min}_{Q}+s-\frac{1}{3}\right)=s-c^{2}_{s{\rm ,~nonderiv},~Q}.
\end{equation}

Now, {it is worth to stress that unlike Eqs. (\ref{eq:deltaMix})--(\ref{eq:csMix}) for the quark-hadron interface,} Eqs. (\ref{eq:deltaQM})--(\ref{eq:csQM}) are {fully} physical and one can extract relevant information (to be explicitly given in the next section) from both of them. In particular, Eq. (\ref{eq:deltaQM}) for $\Delta_{Q}$ show us that if $s>1/3$, then {there} will be a steep decreasing behavior easily approaching negative values in the ultra-dense sector of twin stars {are dynamically stable}. We will see later that this behavior is also manifested in the $\Delta_{Q}=\Delta_{Q}(\mu_{B})$ plane. On the other hand, Eq. (\ref{eq:csQM}) tell us that the derivative and nonderivative terms are complementary if `$s$' is large ($\sim 1$), i.e. increasing $c^{2}_{s{\rm ,nonderiv},Q}$ implies decreasing $c^{2}_{s{\rm ,deriv},Q}$ meaning that the high-negative slope of $\Delta (\epsilon)$ is favored for large `$s$' at increasing energy densities.

\begin{figure*}[!t]
\vspace*{-0.55cm}
\hspace*{-0.65cm}
  \includegraphics[width=0.512\textwidth]{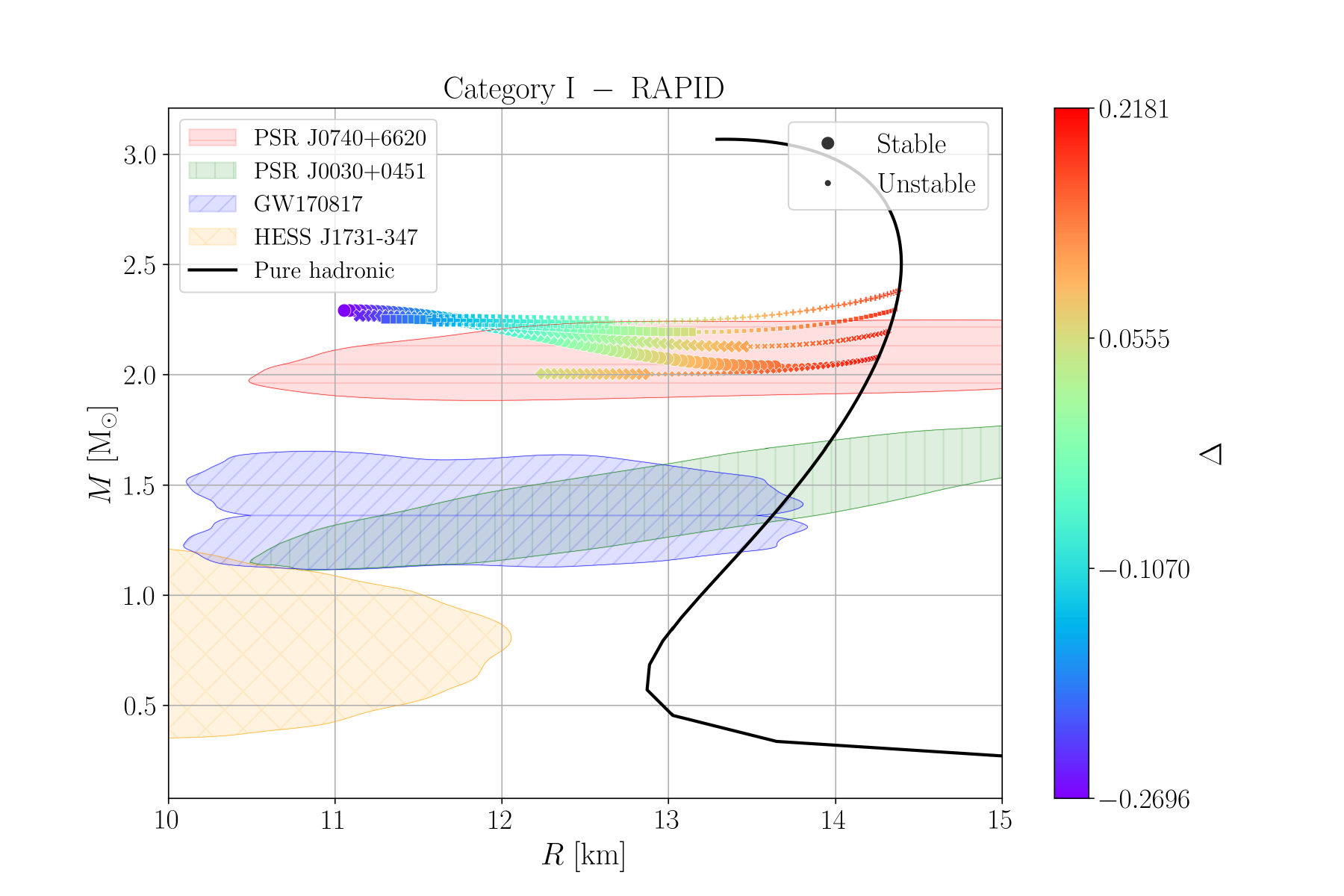}
  \includegraphics[width=0.512\textwidth]{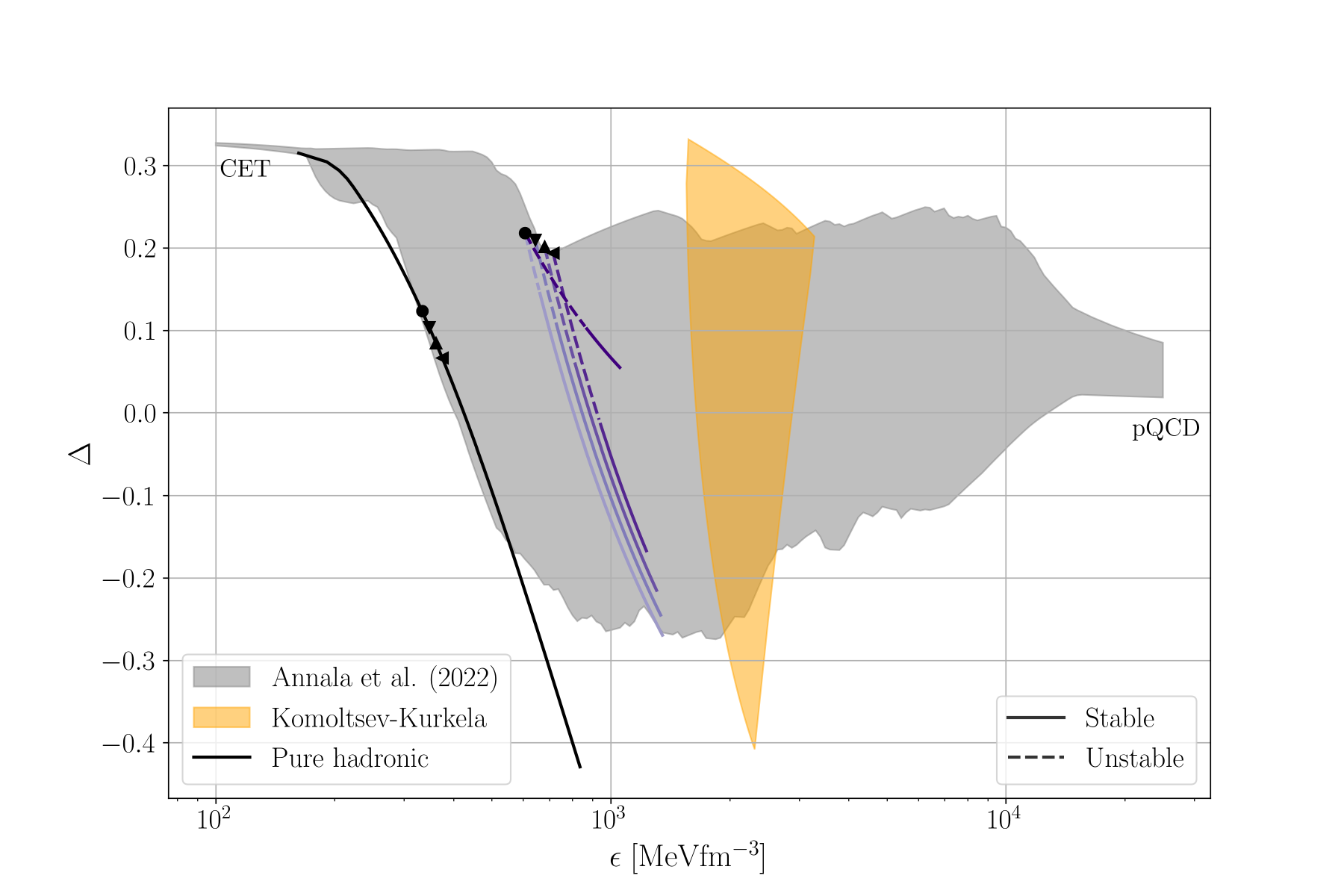}
\hspace*{-0.65cm}
  \includegraphics[width=0.512\textwidth]{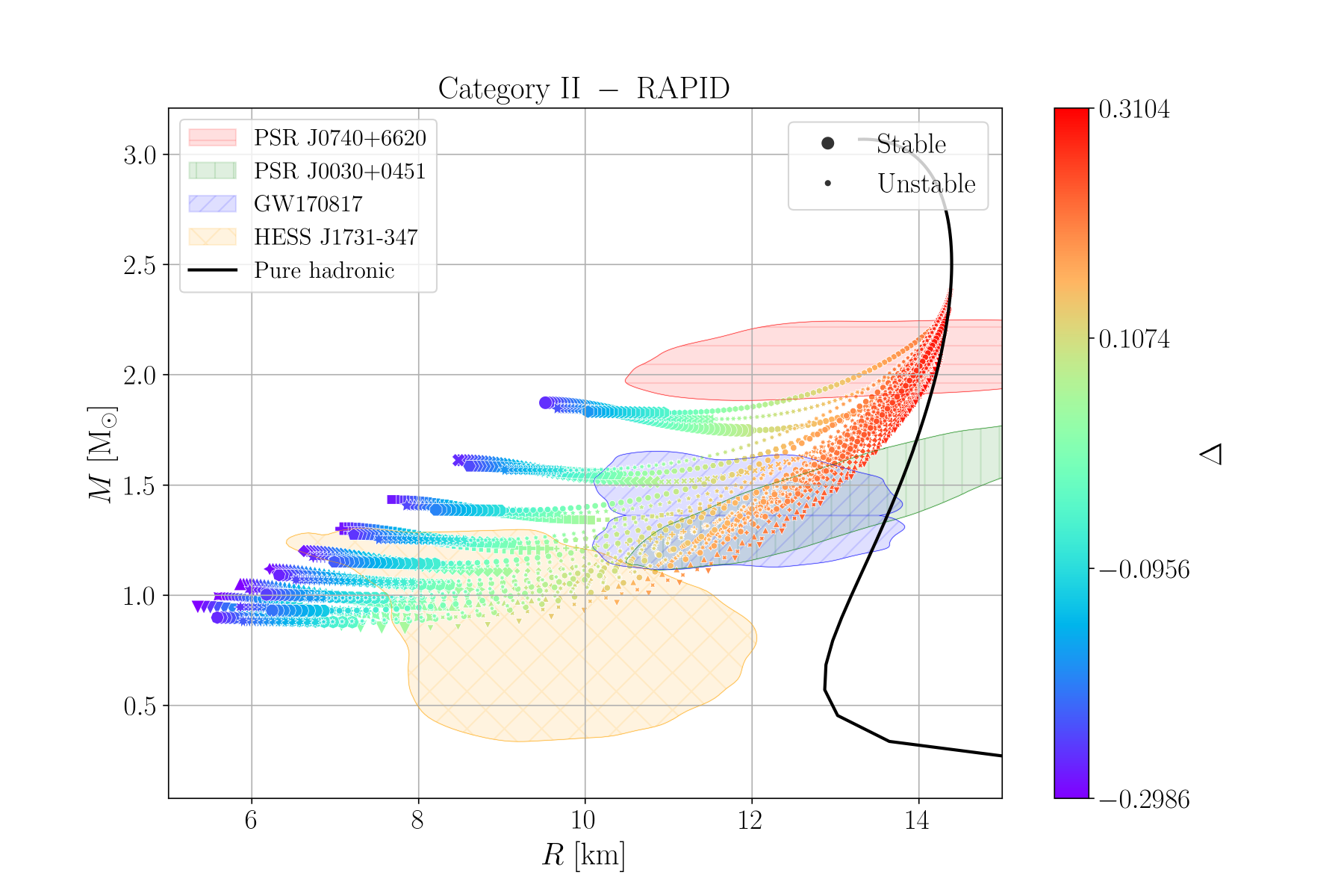}
  \includegraphics[width=0.512\textwidth]{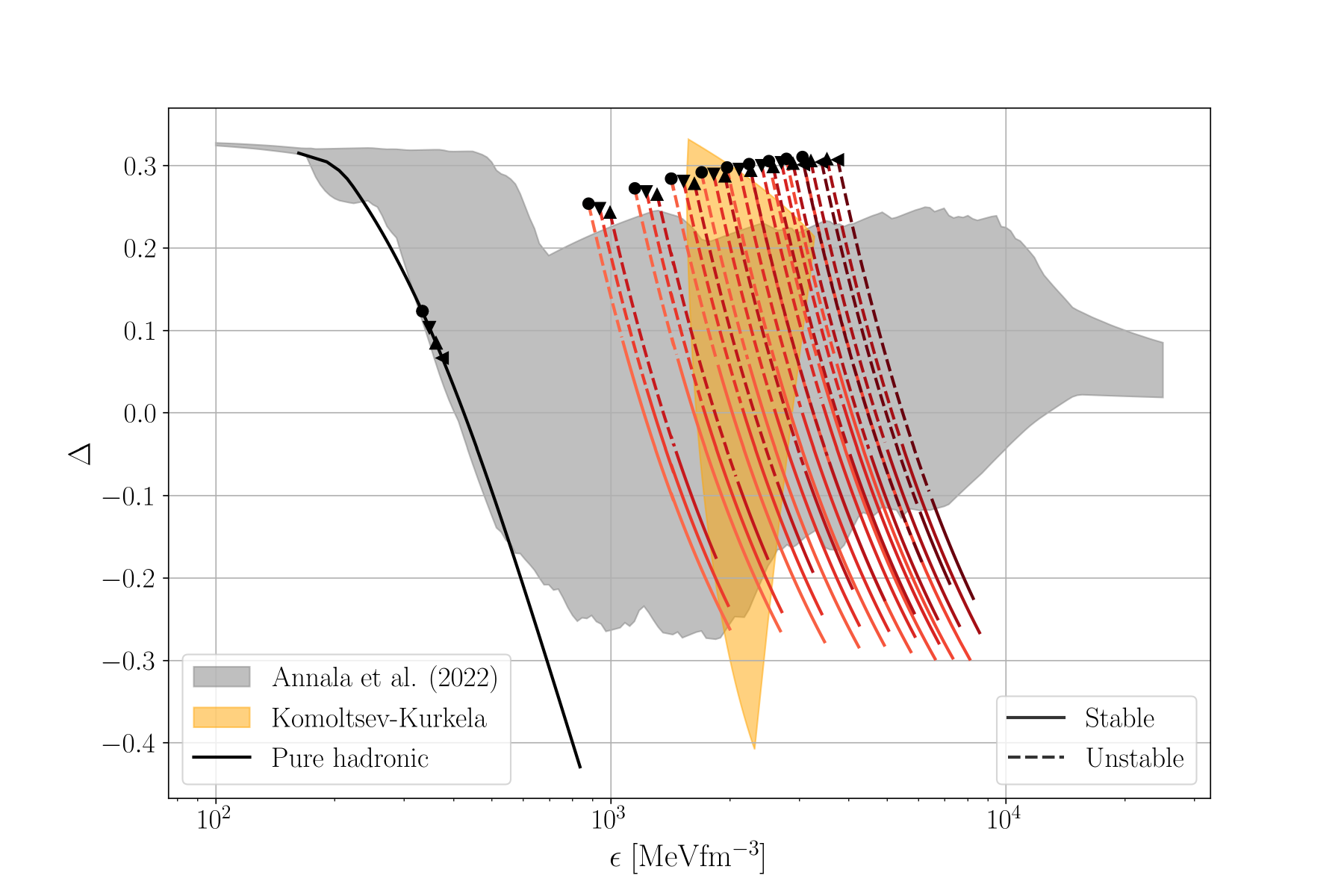}
  \caption{\label{fig:catI_II}
    \textit{Left panels}: $M$--$R$ relations for  twin stars belonging to Categories I (upper panel) and II (lower panel) (including various current astrophysics constraints) with a range of colours on the right side indicating values for their corresponding trace anomalies. Note that thick (thin) colored and filled circles indicate stable (unstable) twins in the sense of radial pulsations.
    \textit{Right panels}: Dense trace anomaly as a function of energy density, $\Delta=\Delta(\epsilon)$, where we include constraints coming from CET (continuous black curve),  Komoltsev-Kurkela~\cite{Komoltsev:2021jzg} (orange band), Annala et al.~\cite{Annala:2021gom} (grey band) and pQCD~\cite{Kurkela:2009gj}. Besides, the black dots and differently-oriented triangles represent the jumps announced in Eq. (\ref{eq:deltaMix}) which one has to read from left to right, e.g. a given left triangle (low energy densities) starts to obey Eq. (\ref{eq:deltaMix}) until the same triangle appears on the right (higher energy densities), where the QM phase begins. Note that (dis)continuous colored QM curves indicate (un)stable twin stars.}
\end{figure*}

\begin{figure*}[!t]
\vspace*{-0.55cm}
\hspace*{-0.65cm}
  \includegraphics[width=0.512\textwidth]{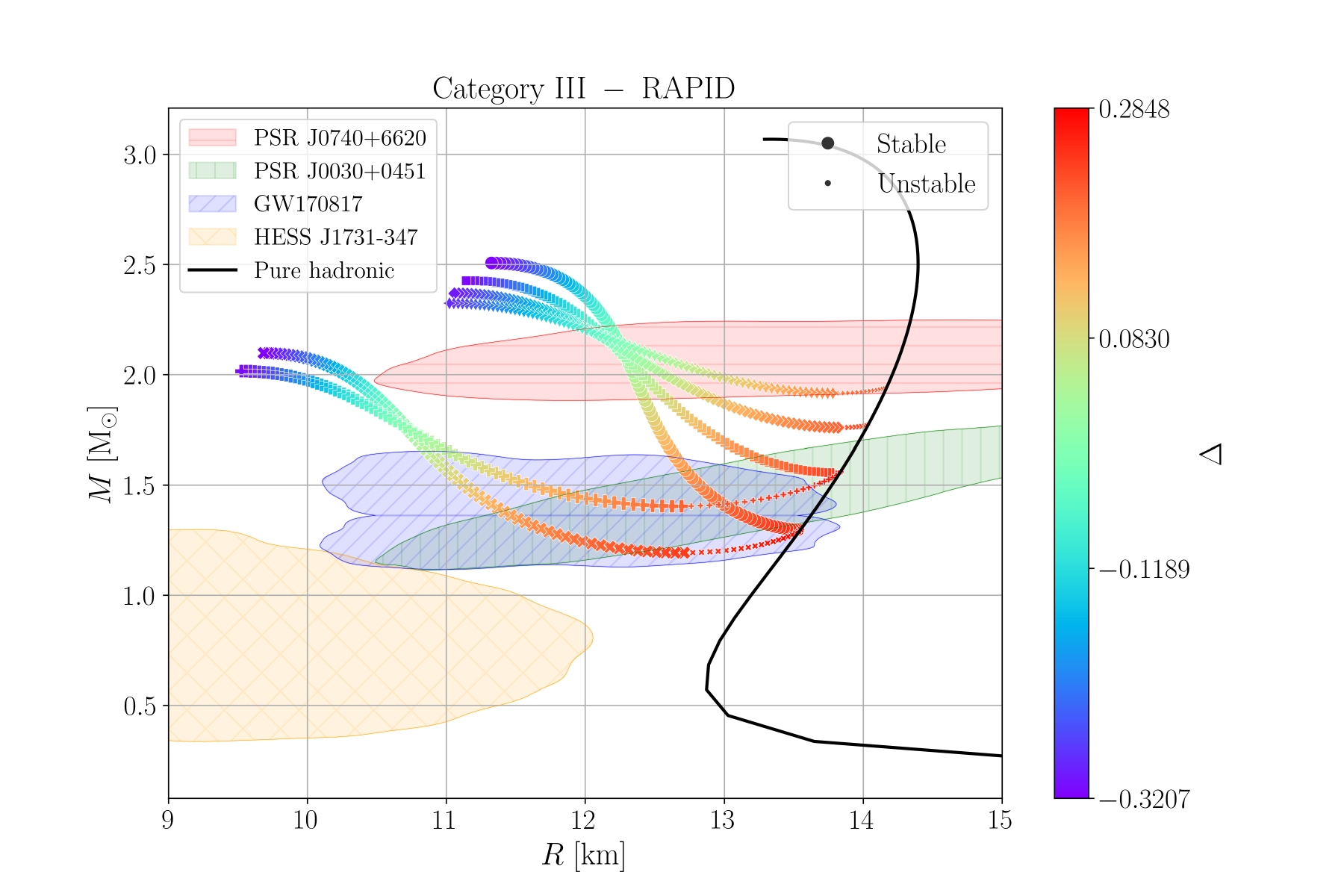}
  \includegraphics[width=0.512\textwidth]{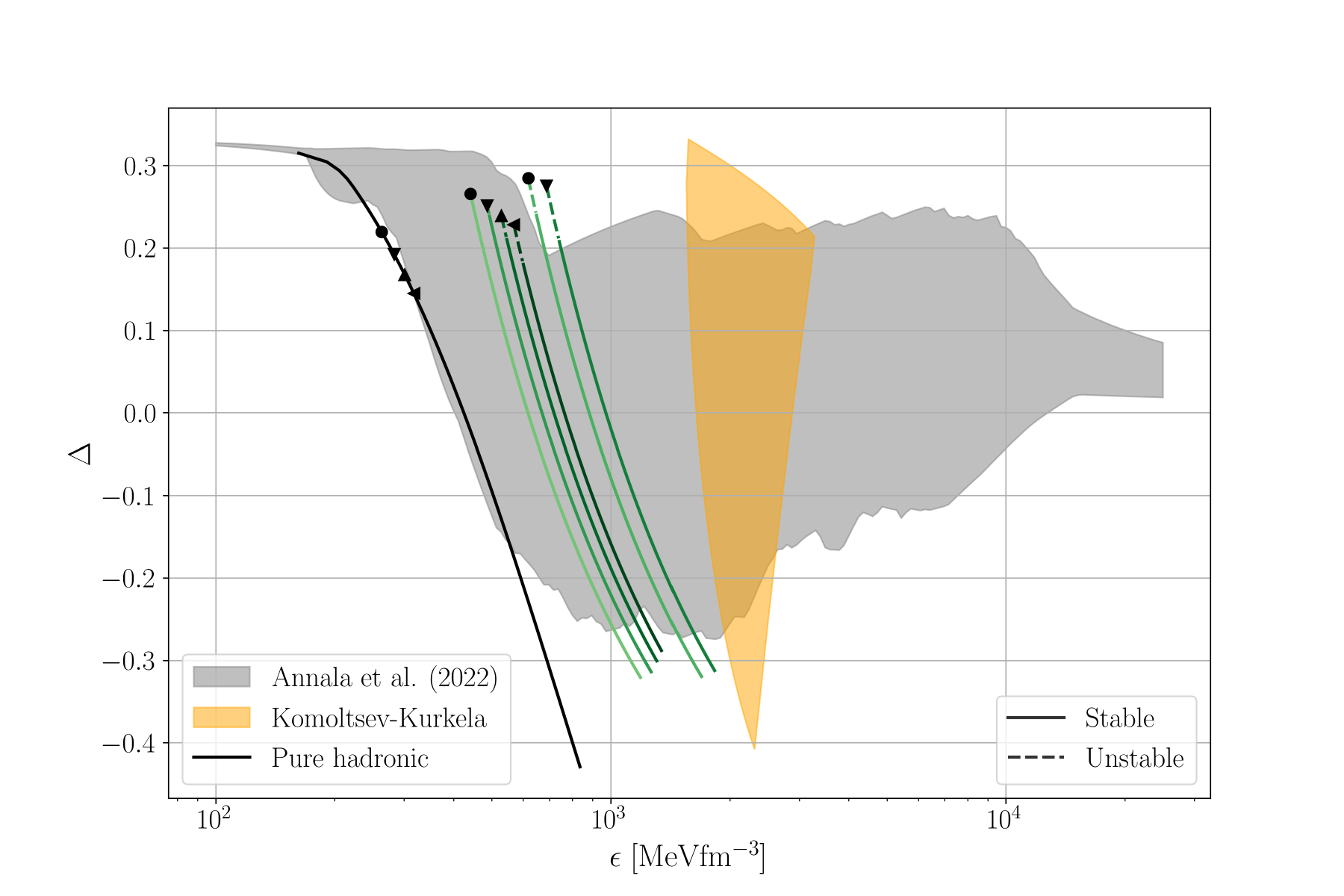}
\hspace*{-0.65cm}
  \includegraphics[width=0.512\textwidth]{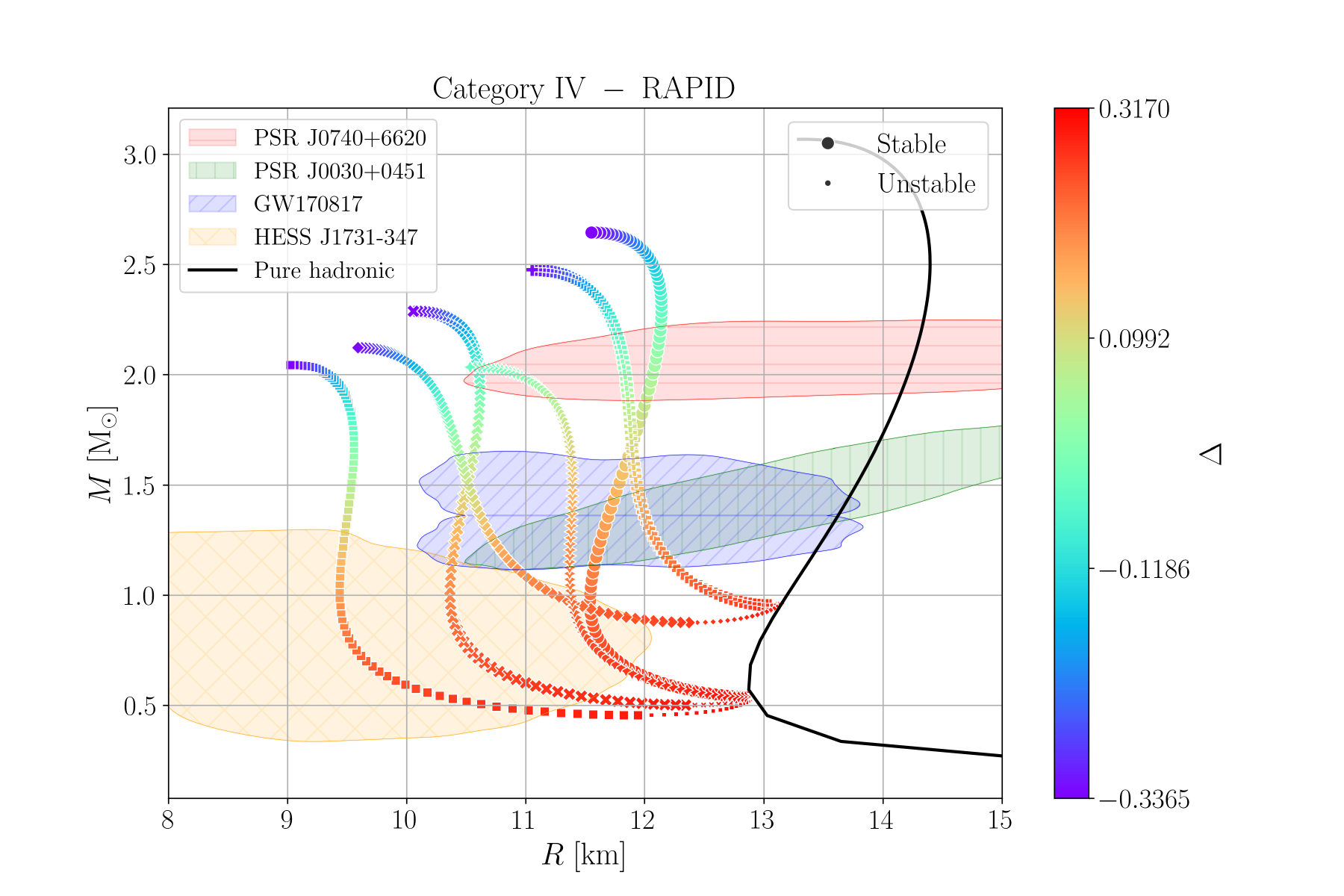}
  \includegraphics[width=0.512\textwidth]{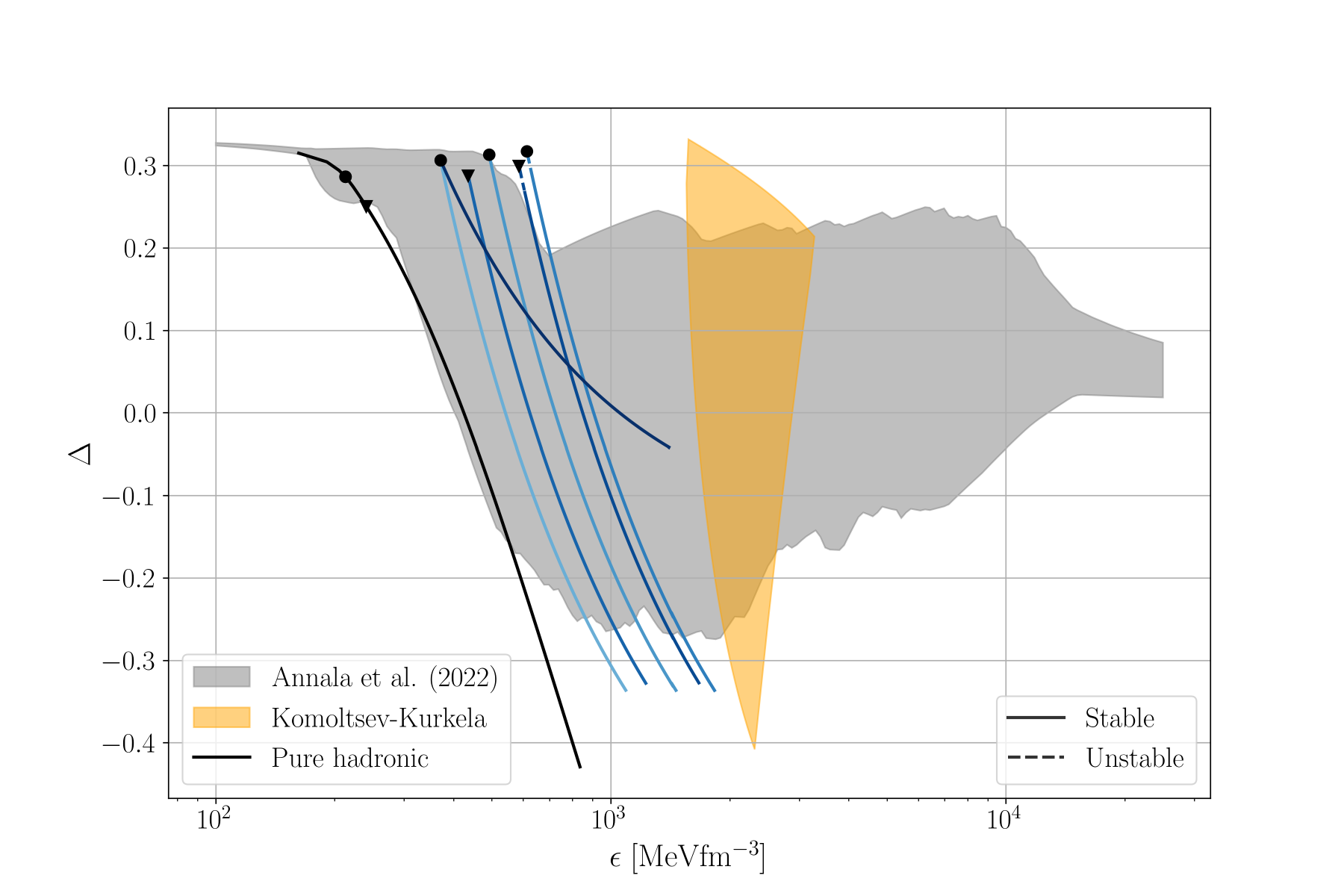}
  \caption{\label{fig:catIII_IV} Same captions and constraints as Fig.~\ref{fig:catI_II} but now for Categories III and IV twin stars.}
\end{figure*}

\subsection*{{How to probe $\Delta$ in twin stars?}} 

{As explained in detail in this and the last section, a careful analysis of the behavior of $\Delta$ at the core of twin stars implies to probe robustly the associated EoSs. In turn, this forces us to look for the central-energy-density regions yielding dynamically stable twin stars for the rapid and slow junction conditions. This allows us to discriminate values of $\Delta$ physically consistent through their realization in nature as potentially observable stellar compact objects. In this sense, we will get robust features of this normalized trace anomaly dictated by the stability of the corresponding twin stars. In particular, we will see that for the slow junction conditions one gets a larger parameter space allowing larger and sizeable negative values of $\Delta$, as we show in the next section.}

\section{Twin-Star Trace Anomalies}
\label{sec:TSanom_I}

We pass {to introduce all our} twin EoSs {in the TOV equations and solve them together with the Gondek's} radial oscillation equations with rapid (Sec. IVA) and slow (Sec. IVB) junction conditions to obtain the whole behavior of $\Delta$ for the chosen parameters of Secs. \ref{sec:twinmatter} and \ref{sec:Tanomaly}. It is worth to mention again that from all the available parameter space satisfying Eq. (\ref{eq:Seidov}) {(yielding thousands of twin stars)}, we only display our findings for tens of EoSs ({for the four} categories) for a clear analysis and {due to} space limitations.

For comparison with all of our figures of this section {(including Fig. \ref{fig:intermediate} of Appendix B)}, we show {in the left panels} the current observational constraints from PSR J0740+6620~\cite{Miller:2019cac}, PSR J0030+0451~\cite{Miller:2019cac}, GW170817 (spectral EOS) \cite{LIGOScientific:2018cki} and HESS J1731-347 \cite{Doroshenko:2022cba}. {Besides, in order to not clutter all our figures}, we have preferred to show only the NICER data from Ref. \cite{Miller:2019cac}, {instead of also adding data from} Ref. \cite{Riley:2019yda}. Moreover, the LIGO data from GW170817 \cite{LIGOScientific:2018cki} {is presented based on the spectral EOS analysis, rather than using the universal relations.}

{On the other hand, in} the right panels {of all our figures in this section (again also in Fig. \ref{fig:intermediate} of Appendix B)}, we put the grey and orange bands of Annala et al.~\cite{Annala:2021gom} and Komoltsev-Kurkela~\cite{Komoltsev:2021jzg}, respectively, {already discussed in} Sec. IIA. {We note that} comparisons {with both of these last works} serve {merely} as a guide {since they did not explore hybrid NS with strong phase transitions, i.e. twin stars in detail. Besides, our parameter space was not adjusted} to agree with theirs.

{With that being said, we present the results as follows: we analyze the four categories of twin stars assuming they satisfy first the rapid-conversion junction conditions to then explore the slow-conversion analogue, both imposed at the hadron/quark interface (see Table \ref{tab:BCs1}). For completeness, we also show in Appendix C the radial profile of $\Delta$, i.e. $\Delta(r)$ for maximum-mass stars of each of the four categories of rapid twin stars.}

\subsection{Rapid conversions}

{\bf Categories I and II:} We present in the left panels of Fig.~\ref{fig:catI_II} the $M$--$R$ relations for twin stars of Categories I and II. {From both panels} one can see that twin stars from those categories satisfy almost all the aforementioned current $M$--$R$ constraints except Category I that marginally respects GW170817 but disagrees with HESS J1731-347. 

Now, in the right panels of Fig.~\ref{fig:catI_II}, we display the corresponding behavior of the twin-star trace anomalies as a function of their energy densities for the same two categories. One can easily see that most configurations not only present $\Delta <0$ for a wide range of densities in the QM phase, but they also decrease abruptly, reaching quickly negative values for ultra-dense {dynamically-stable} twins (continuous curves){, while unstable twin-stars exist for $\Delta>0$.} It is also worth to mention that {the EoSs producing} Category I twins do not approach pQCD since they do not pass through the Komoltsev-Kurkela orange band. {Now, although a similar situation occurs for $\Delta$ in Category II twins about their steep decreasing behavior, $\Delta$ is positive when these twin-stars are dynamically unstable but negative for stable ones. In this sense, realistic (stable) twin-star families require necessarily negative $\Delta$. Besides, the associated twin EoSs pass through the orange band, thus potentially approaching pQCD if extrapolated to ultrahigh densities.} (see discussion in Sec. IIA about the Komoltsev-Kurkela region). 

Besides, we verified that all our twin configurations of Category I with\footnote{{From this point on, we omit the variable `$s$' to connect the explained findings of this section with our discussion of the next one for the CSS parametrization of QM.}} $c^{2}_{s, Q} = 1$ in the QM phase have negative $\Delta$ in the high density regime. {By lowering the value of this free parameter to} $c^{2}_{s, Q} = 0.5$, we were able to find dynamically-stable twins {although} $\Delta$ remaining positive for all densities. One can look at these configurations in the $M$--$R$ diagram and $\Delta=\Delta(\epsilon)$ plane in the upper panels in Fig. \ref{fig:catI_II}, being the hybrid branch with smallest masses and less steep slope, respectively. Besides, unlike Category I, Category II twin stars were only stable for $c^{2}_{s, Q}=1$, leading stringently to negative trace anomalies for high densities. In this sense, considering a high enough central density, stable Category II twin stars always presented negative values of $\Delta$ at their cores.

\begin{figure*}
	\vspace*{-0.55cm}
  \hspace*{-0.67cm}
  \includegraphics[width=0.54\textwidth]{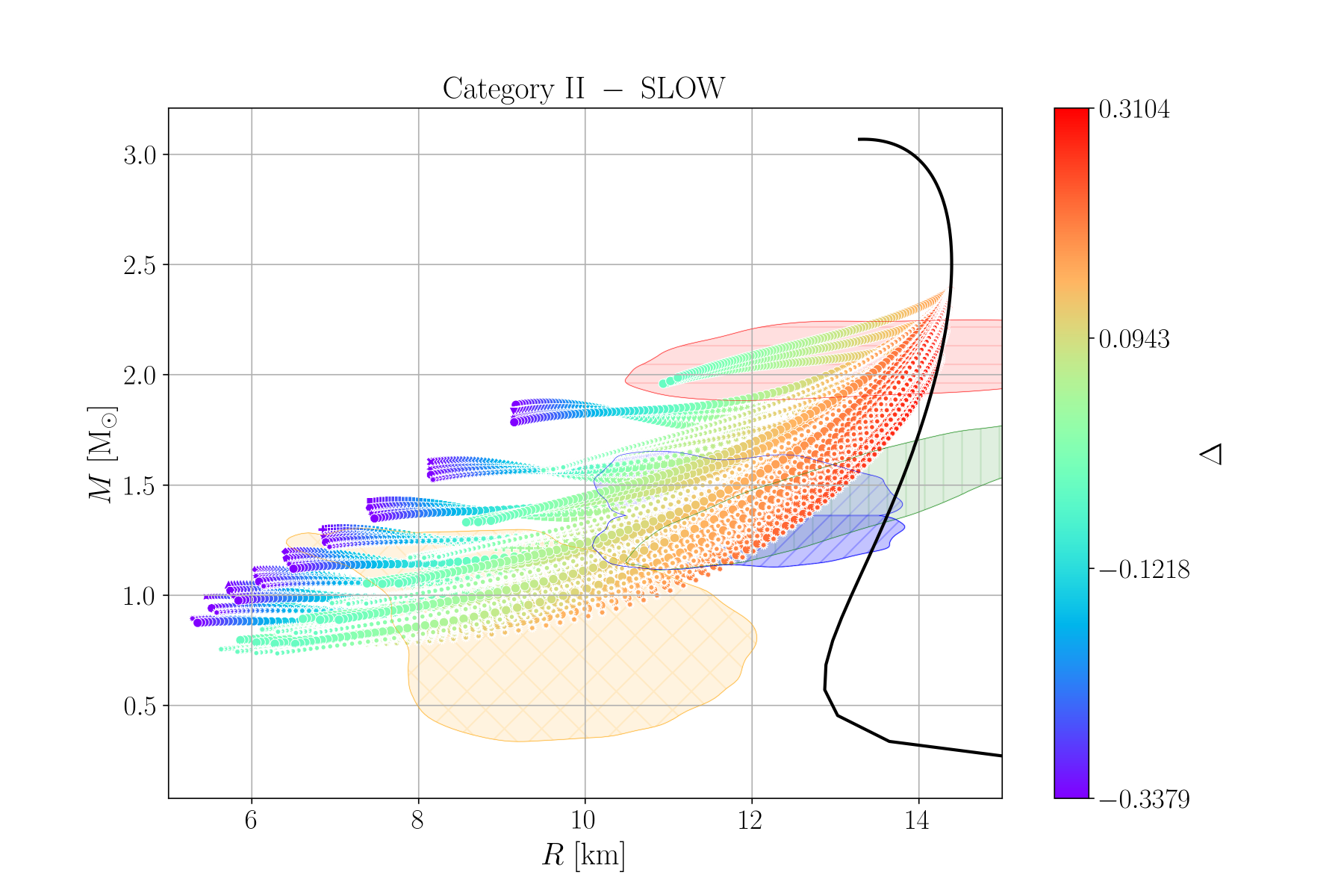}
  \hspace*{-1.1cm}
  \includegraphics[width=0.54\textwidth]{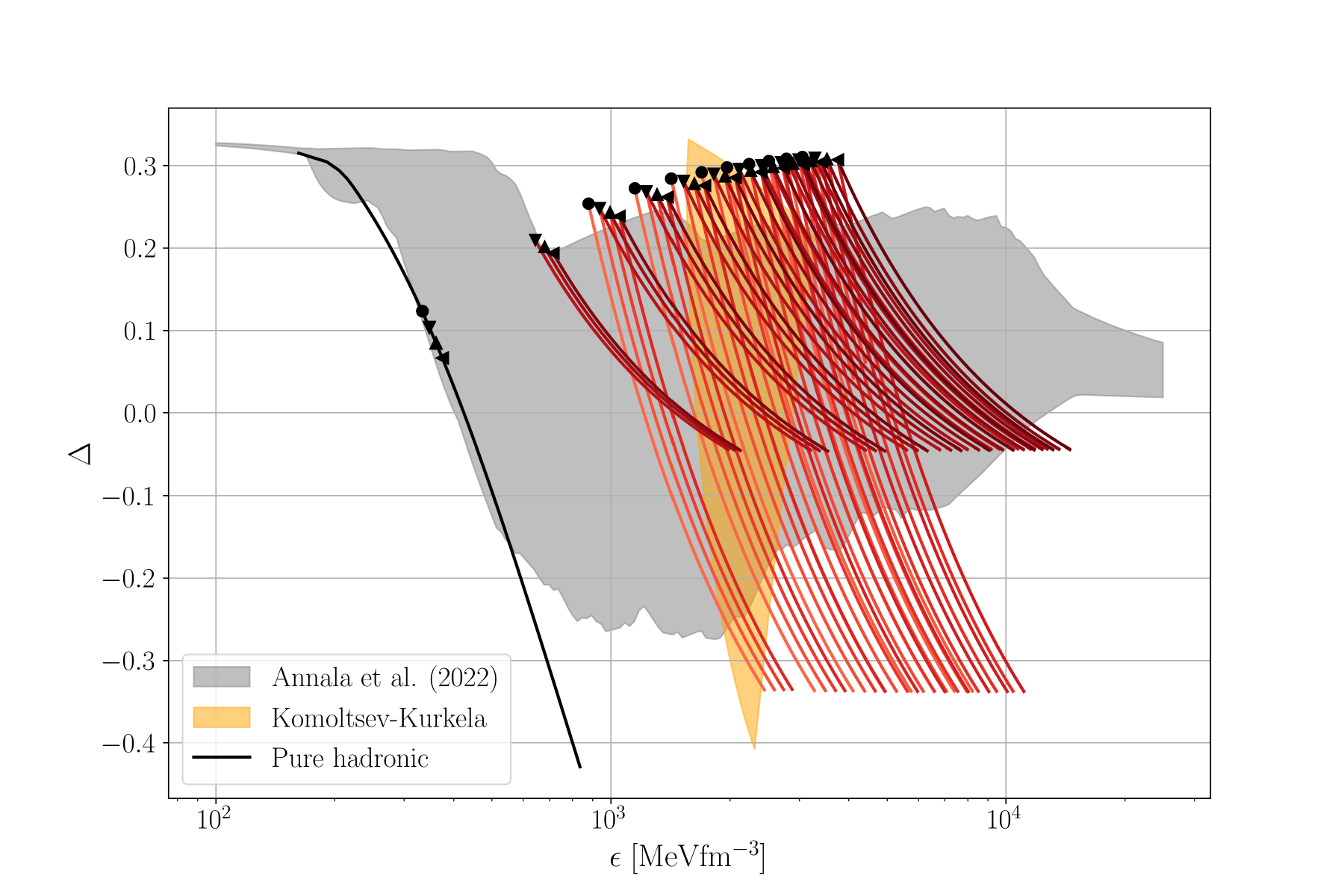}
  \caption{\label{fig:cat_II_slow} Same captions and constraints as in Fig. \ref{fig:catI_II} but for Category II twin stars considering now slow junction conditions.}
\end{figure*}

{\bf Categories III and IV:} {Results for these categories are shown in Fig.~\ref{fig:catIII_IV}. We found similar outcomes that those displayed in the left panels of Fig. \ref{fig:catI_II} for the $M$--$R$ relations (sometimes being in agreement with astrophysics constraints). On the other hand, one can identify in the right panels of Fig.~\ref{fig:catIII_IV} that $\Delta$ is always negative for these massive twin-stars categories, i.e. configurations surpassing the $2M_{\odot}$ limit.}

{It is worth to mention that in both of these categories there is a very small set of unstable twin stars. In other words, almost all of these twin stars in a given $M$-$R$ relation are dynamically stable against radial pulsations. This occurs basically because we fixed the speed of sound of the quark phase at the causality limit, i.e. $c^{2}_{s, Q} = 1$. However, if one instead employs $c^{2}_{s, Q} = 0.5$ in, e.g., Category IV twins, we found a single stable configuration with almost $\Delta \gtrsim 0$ (see the hybrid black curve with lower slope in the right lower panel of Fig.~\ref{fig:catIII_IV}). In this sense, the robustness of the negative values of $\Delta$ in stable twin stars depend strongly on the high values we employ for the (squared) speed of sound for the QM phase which for this work we probed in the range $0.5 \leq c^{2}_{s, Q} \leq 1$.}

{Finally, the reader can realize that the orange region in the right panels of Fig.~\ref{fig:catIII_IV} of the Komoltsev-Kurkela constraint is not reached by the EoSs, i.e. their baryon density sector lies below $n_{B}=10n_{0}$. This implies that these Category III and IV twins do not approach pQCD at high densities. Besides, and as already discussed in Sec. IIA, the gray band of Annala et al.~\cite{Annala:2021gom}  should serve just as a guide when comparing our findings for these categories since their study assumed potential hybrid NSs with smooth transitions, e.g. hadron-quark crossovers.} 

{\bf Twin-star stability through the reaction modes:} In order to close our discussion on these rapid-conversion twin stars, we give some comments connected to the so-called {\it radial reaction modes} \cite{Pereira:2017rmp}. As shown in Ref.~\cite{Pereira:2017rmp}, when one considers rapid junction conditions at the strong-transition point of hybrid stars, it will inevitably lead to the appearance of a reaction mode which, in principle, may be any excited {($n> 0$) radial mode, but it turns out important for the stellar dynamical stability when it becomes} the fundamental one ($n=0$). From our calculations for rapid twin stars, we found that the fundamental mode is always the reaction mode since $\Delta\epsilon$ is a relatively large value, something which is not necessarily true for connected hybrid stars \cite{Alford:2013aca}{, i.e. the hadronic and hybrid branches are joined in the $M$-$R$ diagram without the presence of an unstable branch, as required by the Seidov's criterium. In fact, it is known that in most cases these connected hybrid NSs are dynamically unstable}, as already explored in Ref.~\cite{Pereira:2017rmp}. {In contrast, we will see below that slow-conversion twin stars allow a larger number of stellar configurations due to its effect on the unstable branch between the hadronic and twin star branches, i.e. the slow JC turn this unstable branch into an stable one, something which is impossible with rapid JC.}

\subsection{Slow conversions}

We now move to describe the results of our calculations for {$\Delta$ through the analysis of dynamically-stable} twin star configurations satisfying the {\it slow} junction conditions in the radial pulsation equations, as formulated in Table \ref{tab:BCs2}. We present the $M$--$R$ relations in the left panel of Fig. \ref{fig:cat_II_slow}, while in the right panel its corresponding trace anomalies as functions of energy density.

The main feature of {our obtained} stars is that the hybrid branch becomes connected to the hadronic one in the $M$--$R$ diagram, i.e. both branches remain stable without intermediate unstable stars. {This happens because $\omega^{2}_{n=0}$ remains positive in the unstable branch according to the} the usual instability criterion{, i.e.} $\partial M / \partial \epsilon_{c} < 0$ \cite{Pereira:2017rmp} (see e.g. Ref. \cite{Lugones:2021bkm} for a similar discussion). {In turn, this implies that other energy-density regions in the EoSs are allowed which affects the values of $\Delta$}. 

Besides, many other EoSs {are allowed thus increasing the associated parameter space producing a novel behavior of $\Delta$, i.e. being sizably negative}. {In particular, we found (see Fig.~\ref{fig:cat_II_slow})} many more viable EoSs for Category II twin stars having a stable hybrid branch in the slow scenario. The results for other categories do not change significantly {compared to the rapid junction-condition case although for our present slow case, the stability windows is slightly larger than that obtained from the parameter space satisfying the Seidov's criterium (being in agreement with the rapid case)}, i. e. the last stable configuration occurs for larger central densities when compared to the rapid case. Related to this, {one can also realize that the data of} HESS J1731-347 \cite{Doroshenko:2022cba} {adjusts naturally as a} slow-stable hybrid star~\cite{Mariani:2024gqi}. 

Notice also that in the right panel of Fig.~\ref{fig:cat_II_slow} the curves for {$\Delta$ allowed by the dynamical-stability analysis of these} ultradense twins pass mostly over the Komoltsev-Kurkela orange region although with different slopes, i.e. the darker curves having lower slopes than the lighter ones. {These occurs since the} darker curves have $c_{s,Q}^2 = 0.5$. Moreover, they are present in a significantly larger number of stable configurations than those of the rapid-conversion scenario. In fact, we only found two stable configurations in the rapid case for $c_{s,Q}^2 = 0.5$, for Categories I and IV, with the former presenting an always positive trace anomaly. This does not happen in the slow-conversion scenario where $\Delta < 0$ for higher densities. Nevertheless, there still remain some high-density stellar configurations which accumulate slightly below the pQCD limit at $n_{B}=40n_{0}$ (see discussion of Sec. IIA). 

\section{Discussion}
\label{sec:disc}

From our above results for the discontinuous and {steeply decreasing} twin-star trace anomalies{, $\Delta(\epsilon)$, reaching sizeable negative values}, an immediate question would be to know the behavior $\Delta(\mu_{B})$ {from a full knowledge of $\epsilon(\mu_{B})$ and $P(\mu_{B})$}. {This is relevant since} it could be obtained directly from the QCD thermodynamic Landau potential, $\Omega(\mu_{B})=-P(\mu_{B})$, {analogously to} LQCD at pure $T\neq 0$ \cite{Bazavov:2009zn,Borsanyi:2013bia}. {In fact}, futuristic baryonic results from LQCD (or other novel non-perturbative QCD techniques) could reliably characterize $\Delta$ at $\mu_{B}\neq 0$ and determine if it is truly negative {and steeply} discontinuous around a potential 1st-order transition point. {Today this is still an open question that could rule out twin stars.}

\subsection{{Normalized QCD pressure of twin-star matter}}

For these reasons, it becomes relevant first to understand the behavior of $P=P(\mu_{B})$, or more precisely, its normalized form, $P/P_{\rm SB}$, with $P_{\rm SB}=(3/4\pi^{2})(\mu_{B}/3)^{4}$ as the Stefan-Boltzmann pressure. {It should kept in mind that `$P$' here refers a matching between the CET hadronic EoS at low densities to the CSS EoS for quark matter at intermediate and high densities.} In fact, within the CSS parametrization, it was proven in Ref. \cite{Alford:2013aca} that the quark pressure, $P_{Q}$, could be written as
\begin{equation}
\label{eq:CSSpress}
P_{Q}(\mu_{B})=N\mu^{1+\gamma}_{B}-B.
\end{equation}
{We list in} Table \ref{tab:Q_parameters} the values chosen for these CSS parameters for a representative twin star of each category.

\begin{figure*}
	\vspace*{-0.47cm}
  \hspace*{-0.65cm}
  \includegraphics[width=0.54\textwidth]{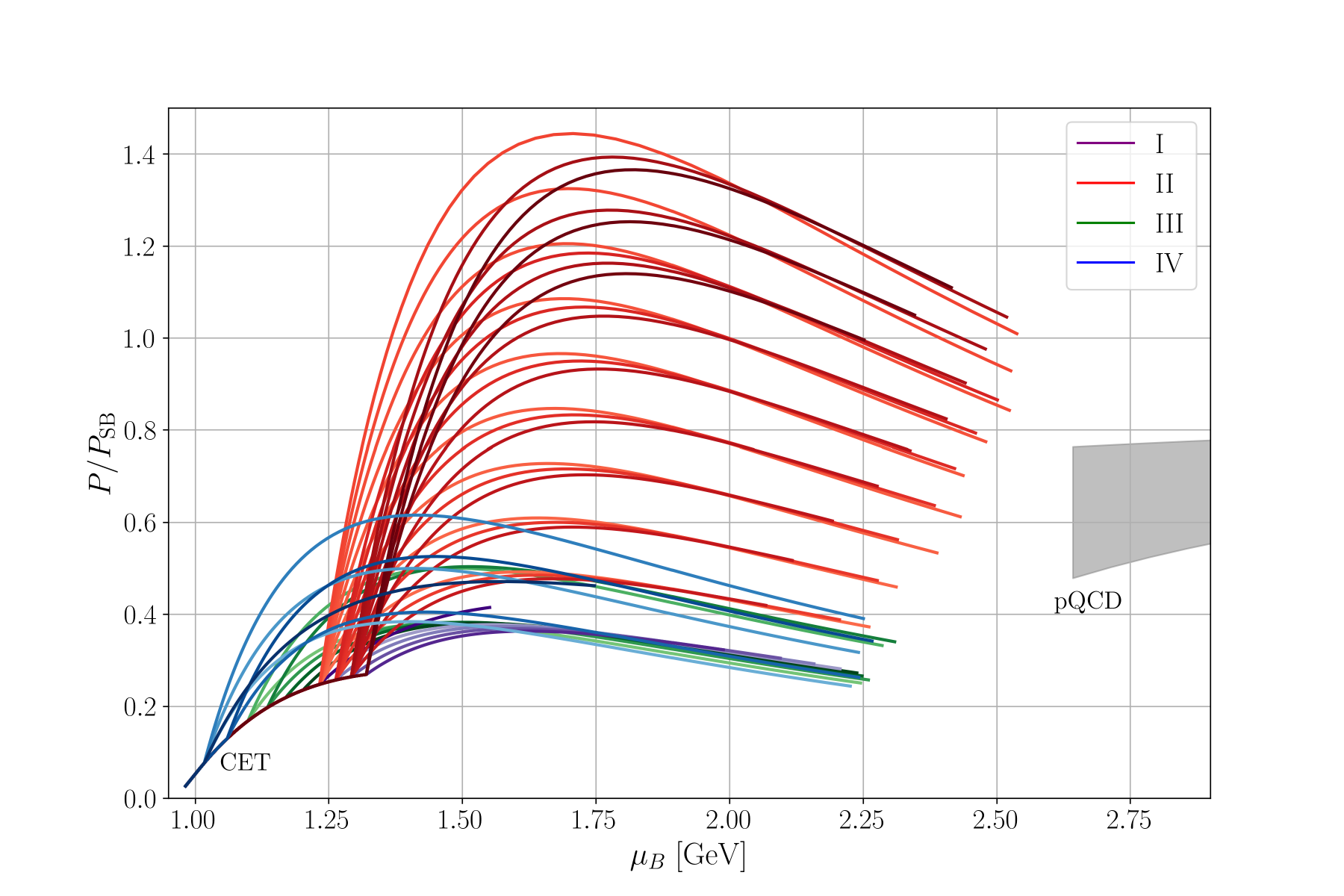}
  \hspace*{-1.1cm}
  \includegraphics[width=0.54\textwidth]{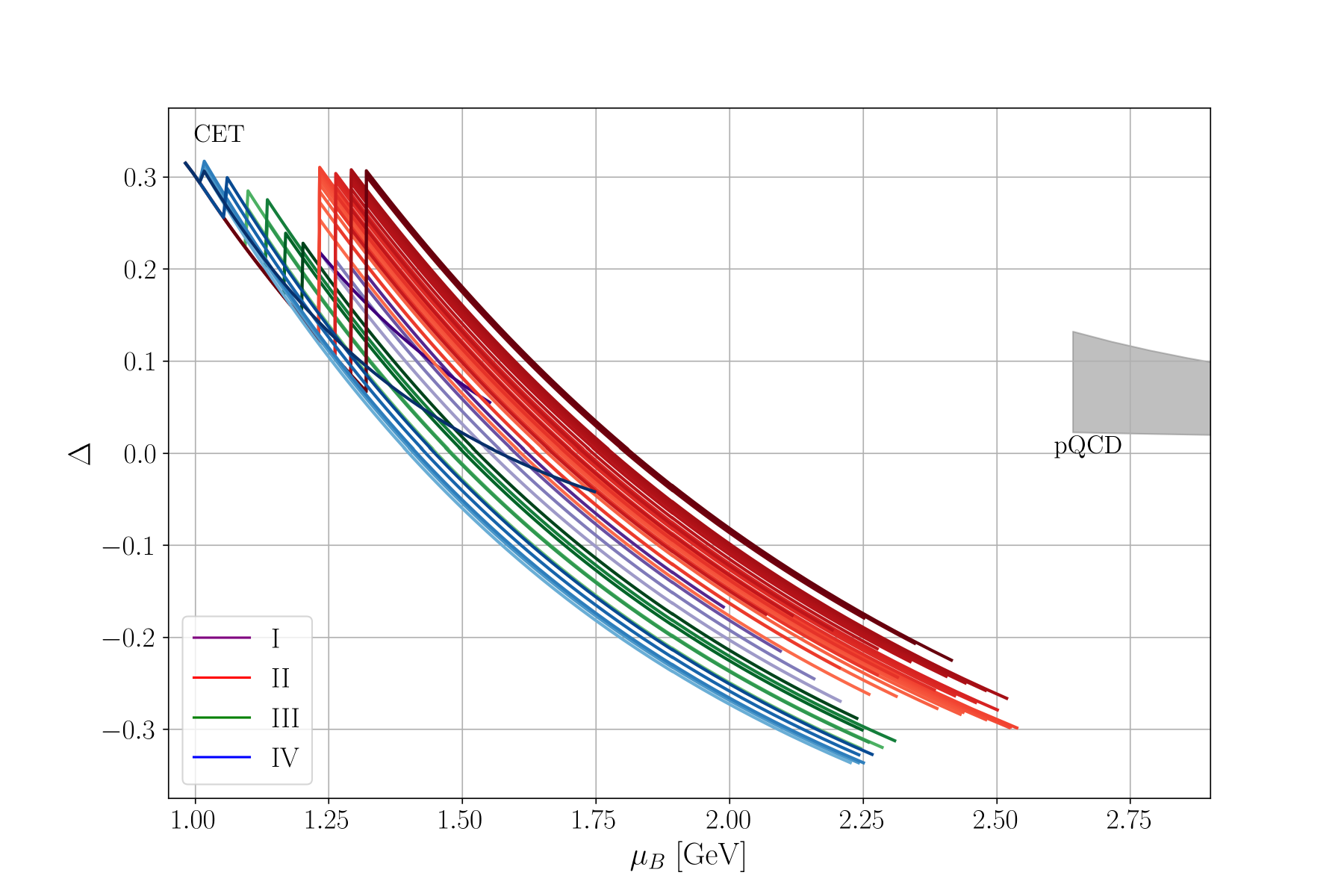}
  \caption{\label{fig:PvsmuB} \textit{Left panel:} Behavior of our {\it rapid} twin-star pressures (normalized by a Stefan-Boltzmann gas) for each category vs baryon chemical potential. Notice the presence of the hadronic CET and quark pQCD bands, this last one appearing at $\mu_{B}=2.6$ GeV corresponding to $n_{B}=40n_0$, as in other works (see e.g. Ref. \cite{Annala:2021gom}). \textit{Right panel:} Dense trace anomalies vs baryochemical potential for the four categories of twin stars. Besides, the pQCD band ($\mu_{B}\geq 2.6$ GeV) is thicker than of CET.}
\end{figure*}

{Moreover,} we display (for the first time in the literature of twin stars with the CSS parametrization) in Fig.~\ref{fig:PvsmuB} our findings for $P/P_{\rm SB}$ and $\Delta$ vs $\mu_{B}$ for the same twin-star EoSs shown in the left panel of Fig. \ref{fig:stiff_rapid_eos}. {In particular, one can see that the quark sector, i.e. $P_{Q}/P_{\rm SB}$, displays an increasing behavior from the transition point (critical baryochemical potential), reaches a maximum, and then begins to moderately decrease. However, our choices of $c^{2}_{s,Q}=\left\lbrace 0.5,1 \right\rbrace$ will surpass the conformal limit of $P/P_{\rm SB}=1$ or, equivalently, $c^{2}_{s,Q}=1/3$. Thus, the previously mentioned maxima in $P/P_{\rm SB}$ will go above the conformal bound for the quark phase in twin stars.}

{Being more specific,} Category III and IV twin stars do not change too abruptly their normalized pressure at the transition point. Moreover, they reach maximal values of $P/P_{\rm SB} \sim 0.4\pm 0.2$, from which it seems difficult for those pressures to reach the pQCD (grey) band at $\mu_{B}=2.6$ GeV. However, Category I and II twin stars present strong discontinuities with maximal values reaching even $P/P_{\rm SB}\sim 1 \pm 0.4$, thus showing a favored trend towards pQCD if extrapolated to higher densities. 

Some comments about the relation of these last findings compared to current studies of the NS EoS are in order. In particular, Bayesian approaches (see, e.g. \cite{Annala:2023cwx,Gorda:2023usm,Providencia:2023rxc}) and other works having microphysics control (see, e.g. Ref. \cite{Jakobus:2020nxw}) sometimes build their predictions for the EoS at intermediate densities not only through relations like $P=P(\epsilon)$, but also in the plane $P/P_{\rm SB}$ vs $\mu_{B}$. Remarkably, most of these works always assume bands of uncertainty satisfying $P/P_{\rm SB}\leq1$ {(even considering the potential presence of phase transitions) in order to reach pQCD at high densities by increasing the values of $P/P_{\rm SB}$ but, more importantly, from below the pQCD band.} However, our results of the left panel of Fig. \ref{fig:PvsmuB} for Category I and II twin stars indicates that it could still possible to reach pQCD with a decreasing $P/P_{\rm SB}$, i.e. from above {with maximal values of} $[P/P_{\rm SB}]_{\rm max}$ as large as $1.4$. Thus, this novel and intriguing possibility could further constrain the NS EoS {in, e.g., Bayesian studies}.

\subsection{{Transition degrees of freedom \\and latent heat in twin stars}}

Furthermore, a few years ago, Ref. \cite{Fujimoto:2022ohj} pointed out that $\Delta \geq 0$ in NS interiors due to the increasing number of effective degrees of freedom, $N_{\rm eff}$, i.e. $\Delta \sim d(P/P_{\rm SB})/d\mu_{B}\sim dN_{\rm eff}/d\mu_{B}$, {thus being} always positive definite. Our viewpoint about $\Delta(\mu_{B})$ with strong transition is the following: these discontinuous transitions are tightly connected to the latent heat, $Q^{*}$, through

\begin{table}
  \begin{tabular}{c|c|c|c|c|c}
      Category & $Q^{*}$ & $\gamma=1/c^{2}_{s,Q}$ & $\epsilon_0$ & $N$ & $B$ \\ 
      \hline
      I & 274.26 & 1.0 & 537.34 & 2.22$\times 10^{-4}$ & 268.67 \\
      II & 545.80 & 1.0 & 808.88 & 3.12$\times 10^{-4}$ & 404.44 \\
      III & 178.38 & 1.0 & 411.62 & 1.95$\times 10^{-4}$ & 205.81 \\
      IV & 157.94 & 1.0 & 360.65 & 1.84$\times 10^{-4}$ & 180.43
  \end{tabular}
  \caption{\label{tab:Q_parameters} {List of CSS parameters including the} latent heat from Eq.~(\ref{eq:latentheat}) for a representative of each twin-star category. They are required to get $\mu_{B}$ and $n_{B}$ in the CSS parametrization~\cite{Alford:2013aca}. Note that $\left\lbrace Q^{*}, \epsilon_0=(1+\gamma)B,B \right\rbrace$ are in MeV\,fm$^{-3}$, while `$N$' is in MeV$^{-\gamma}$\,fm$^{-3}$ and `$\gamma$' is in units of the inverse of the speed of light squared, i.e. dimensionless in this work.}
\end{table}

\begin{equation}
\label{eq:latentheat}
Q^{*}=\mu_{c}\Delta n_{B}=\left\langle T^{\mu}_{\mu}(\mu_{B}^{+}\to \mu_{c}) \right\rangle_{Q}-\left\langle T^{\mu}_{\mu}(\mu_{B}^{-}\to \mu_{c}) \right\rangle_{H},
\end{equation}
where the symbols $\mu_{B}^{+(-)}\to \mu_{c}$ mean `approach from the left (right) to the critical baryochemical potential $\mu_{c}$' {and $\left\langle ... \right\rangle_{Q (H)}$ refers to the quark (hadron) ensemble average}. It is straight to infer from Table \ref{tab:Q_parameters} that these $Q$'s are non-negligible in these four twin-star categories{, in particular, Categories I and II where $P/P_{\rm SB}>1$ easily.} 

Interestingly, one can also write a normalized `$Q^{*}$' as
\begin{equation}
\label{eq:latentN}
\frac{Q^{*}}{\mu^{4}_{c}}=\mu_{c}\left[\left(\frac{dN^{Q}_{\rm eff}}{d\mu^{+}_{B}}\right)-\left(\frac{dN^{H}_{\rm eff}}{d\mu^{-}_{B}}\right)\right]_{\mu_{B}^{\pm}\to \mu_{c}},
\end{equation}
where $N^{Q}_{\rm eff}\equiv P_{Q}/\mu^{4}_{B}$ and $N^{H}_{\rm eff}\equiv P_{H}/\mu^{4}_{B}$ are the effective degrees of freedom at each quark (Q) and hadronic (H) phases. {Notice that since should satisfy $Q^{*}>0$, then $dN^{Q}_{\rm eff}/d\mu_{c} > dN^{H}_{\rm eff}/d\mu_{c}$, thus favoring not only the increment of degrees of freedom in the quark phase but also its speed at the transition. Since for us $Q^{*}$, it is just a free parameter, we are assuming implicitly large $dN^{Q}_{\rm eff}/d\mu_{c}$ for large chosen $Q^{*}$. This can be seen in the left panel of Fig. \ref{fig:PvsmuB} for Categories I and II, where the almost vertical curves around the transition point represent the large $Q^{*}\sim dN^{Q}_{\rm eff}/d\mu_{c} \sim d(P/P_{\rm SB})/d\mu_{c}$, as shown in Table \ref{tab:css_parameters}.}

{Thus, 1st-order transitions modify abruptly the degrees of freedom in the system through a discontinuous behavior of the trace anomaly, Eq. (\ref{eq:latentheat}), requiring a finite $Q^{*}$. This leads (as already proven in Sec. \ref{sec:TSanom_I}) to non-negligible negative values for $\Delta$. In fact, one can consider this delicate and non-trivial observation as a natural extension of the initial study started by} Ref. \cite{Fujimoto:2022ohj}.

{Now, it is worth to mention that somewhat recently Ref. \cite{Fujimoto:2024ymt} explored (slightly) weak 1st-order transitions in NSs having $Q\sim 0$ since at $\epsilon\sim 700~{\rm MeV fm^{-3}}$ one has $N^{H}_{\rm eff} \sim N^{Q}_{\rm eff}=$constant. From their viewpoint, this implies a saturation of degrees of freedom, which provides a little (negligible) effect on $\Delta$ to be noticed. This is in marked contrast to our findings of Sec. \ref{sec:TSanom_I} for twin stars.}

\subsection{{Normalized QCD trace anomaly of twin stars}}

The corresponding $\Delta(\mu_{B})$ for twin star matter at all densities can be written as follows:
\begin{equation}
  \label{eq:DeltaModel}
  \Delta(\mu_{B}) = \begin{cases}
    \Delta_{H}(\mu_{B}) \hfill \mu_{B} <\mu_{c} \,, \\
    \Delta_{Q}(\mu_{B}) \hspace{.5cm} \mu_{B} > \mu_{c}\, ,
  \end{cases}
\end{equation}
where at $\mu_{B}=\mu_{c}$ it suffers a discontinuous jump. This can be clearly seen {if one uses the CET and CSS EoSs directly. Besides, we show our results for $\Delta$} in the right panel of Fig. \ref{fig:PvsmuB}. One can see that at intermediate `$\mu_{B}$' the four categories decrease until `$\mu_{c}$' (vertical lines). Then, they increase to then decrease discontinuously (and somewhat steeply) in the QM phase to reach sizeable negative values. Furthermore, the latent heats, $Q^{*}$, for Category I and II twin stars are larger compared to Categories III and IV, thus signaling a substantial change of effective degrees of freedom, according to Eq. (\ref{eq:latentN}). 

Noteworthily, the normalized quark trace anomaly, $\Delta_{Q}$, can be explicitly written from Eq. (\ref{eq:CSSpress}) as
\begin{equation}
\label{eq:DeltaQ}
\Delta_{Q}=\frac{4B-{3N\gamma \mu^{1+\gamma}_{B}\left(c^{2}_{s,Q}-{1}/{3}\right)}}{B+N\gamma \mu^{1+\gamma}_{B}},
\end{equation}
being positive (negative) if $c^{2}_{s,Q}<1/3~(>1/3)$, and of course, assuming reasonable values for `$B$' and `$N$', as those listed in Table \ref{tab:Q_parameters}. {In this way, one can see again that large $c^{2}_{s,Q}$ yields very negative $\Delta$ at high densities.}

\subsection{Why `$\Delta<0$' at twin-star cores?}

As somewhat discussed in previous sections, only {reliable} non-perturbative QCD calculations of $\Delta(\mu_{B})$ will {be able to} fully explain its {potential} negativeness {through a steep decreasing behavior above} the transition point {if the QCD deconfinement transition is of first order}. However, while waiting for those results, we can still provide some estimates that exhibit the main features of our numerical findings. {In particular, we pass to employ simplified/analytic models for hadronic and quark matter to} probe {how} $\Delta<0$ emerges naturally in a robust manner. {In fact, we will see that a $\Delta < 0$ depends strongly on the fixed values for the} free parameters {required to get stable twin stars. In this sense,} we elaborate models for $\Delta$ for each phase, i.e. hadronic ($\Delta_{H}$) and quark ($\Delta_{Q}$) matter.

\subsection*{\bf Hadronic model} 

As in Ref. \cite{Fujimoto:2022ohj}, we choose a model inspired by mean-field quantum hadrodynamics with
\begin{equation}
P_{H}=\left(\frac{C^{*}}{\Lambda^{2}} \right) n^{2}_{B}=\left(\frac{\Lambda^{2}}{4C^{*}}\right)(\mu_{B}-m_{N})^{2},
\end{equation}
where $m_{N}=N_{c}\Lambda_{\rm QCD}$ is the baryon mass ($N_{c}$ are the number of colors and $\Lambda_{\rm QCD}$ the non-perturbative QCD scale), $C^{*}$ is the typical interaction strength, and $\Lambda$ the scale of the system. Within this model, one has that the non-normalized trace anomaly {for this hadronic phase} is

\begin{equation}
\left\langle T^{\mu}_{\mu} \right\rangle_{H}=m_{N}n_{B}-2\left(\frac{C^{*}}{\Lambda^{2}}\right)n^{2}_{B},
\end{equation}

which, in principle, can be negative, $\left\langle T^{\mu}_{\mu} \right\rangle_{H} < 0$, iff there is some critical baryon density $n^{*}_{B}$ from which
\begin{equation}
n_{B}>\frac{m_{N}\Lambda^{2}}{2C^{*}}(\equiv n^{*}_{B})~{\rm or}~n_{B}=a~n^{*}_{B},
\end{equation}
being $a>1$. Interestingly, from this one can directly obtain a simple expression for the (negative) normalized trace anomaly as
\begin{equation}
\label{eq:DeltaH}
\Delta_{H}=\frac{2}{3}\left(\frac{1-a}{a+2}\right)<0~{\rm if}~a>1.
\end{equation}
This implies its (squared) speed of sound to be
\begin{equation}
c^{2}_{s,H}=\frac{2C^{*}}{\Lambda^{2}}\left(\frac{n_{B}}{m_{N}+2(C^{*}/\Lambda^{2})n_{B}}\right)=\frac{a}{1+a}.
\end{equation}
{This means that} for the lowest (highest) values satisfying this condition, one has, e.g., $a=1.01~(1000)$ giving $c^{2}_{s,H}\simeq 0.5~(1)$, respectively. This leads us to infer that any hadronic model will have $\Delta_{H}<0$ for large speeds of sound. In fact, this was already found in Ref. \cite{Jimenez:2019iuc} for the well-known hadronic APR and TM1 EoSs having large $c^{2}_{s}$ at intermediate densities. 

Besides, in the right panels of Figs. \ref{fig:catI_II} to \ref{fig:cat_II_slow}, we present results of $\Delta$ for our CET model ({\it pure hadronic}, black curves), where one can easily see their negative behavior. 
{This is in agreement with the  main insight given by our quantum-hadrodynamic model since it is known that CET produce large speeds of sound at intermediate densities \cite{Hebeler:2013nza}. Since this result is relevant, i.e. $\Delta_{H}<0$ for large sound speeds, we prove in Appendix D the robustness of this finding by employing a polytropic model for hadronic matter, where similar features are found.}

\subsection*{Quark model}

Inspired by Ref. \cite{Alford:2013aca} for its CSS EoS and corresponding Eq. (\ref{eq:CSSpress}) for $P_{Q}(\mu_{B})$, we propose the more generic pressure including a condensation term which could potentially lead to the same CSS EoS {for $P_{Q}(\epsilon)$}. This is motivated by the fact that for cold and finite-isospin matter, LQCD calculations \cite{Abbott:2023coj} and effective models \cite{Chiba:2023ftg} have concluded that the main reason for their $\Delta$ be negative is the dominating condensation term at intermediate densities. We will test its importance in this CSS model for intermediate $\mu_{B}$ at the neighborhood of the transition point. Thus, our proposed pressure is\footnote{Electrons have a small contribution, which we neglect for sake of simplicity. Besides, do not confuse our constant `$\gamma$' with $\gamma=d\ln P/d\ln \epsilon$ usually found in the literature, e.g. as in Ref. \cite{Annala:2023cwx}.}
\begin{equation}
\label{eq:new}
P_{Q}\equiv N\mu^{1+\gamma}_{B}+C\mu^{(1+\gamma)/2}_{B}-B,
\end{equation}
where `$N$' and `$C$' are related to `$N_{\rm eff}$' and a non-perturbative {\it condensation} term, respectively. In general, they are running quantities of $\mu_{B}$, i.e. $N=N(\mu_{B})$ and $C=C(\mu_{B})$. Besides, according to this parametrization \cite{Alford:2013aca}, $\gamma \equiv 1/c^{2}_{s,Q}=$ constant and `$B$' would be {a constant resembling the MIT bag (confinement)} constant. {Besides, notice that in the conformal limit with $c^{2}_{s,Q}=1/3$, one has $\gamma=3$ for which the factor multiplying `$C$' has a power `$(1+\gamma)/2=2$'} in order to reproduce known color-flavor locked (CFL) superconductivity results giving $P_{\rm CFL}\sim \mu^{2}_{B}$, which in total gives the widespread \cite{Zhang:2020jmb,VasquezFlores:2010eq,Pereira:2017rmp} pressure $\sim a_{4}\mu^{4}_{B}+a_{2}\mu^{2}_{B}-B$, being $a_{2},a_{4}$ some phenomenological parameters.

As expected, one can also deduce from Eq. (\ref{eq:new}) the associated trace anomaly {for this quark phase:}
\begin{multline}
\label{eq:traceCSSmu}
\left\langle T^{\mu}_{\mu} \right\rangle_{\rm Q} = \left[N(\gamma-3)+\frac{\partial N}{\partial \ln \mu_{B}}\right]\mu^{1+\gamma}_{B}\\
+\left[\left(\frac{\gamma-7}{2}\right) C+\frac{\partial C}{\partial \ln \mu_{B}}\right]\mu^{(1+\gamma)/2}_{B}+4B,
\end{multline}
where we considered the general case of running `$N$' and `$C$'. From Eq. (\ref{eq:traceCSSmu}), one realizes that $\left\langle T^{\mu}_{\mu} \right\rangle_{\rm Q}$ can be positive or negative, depending mainly on `$\gamma$'. For instance, when $\gamma=3~(c^{2}_{s,Q}=1/3)$ and $\left\lbrace N,C \right\rbrace$=~constants, one gets $\left\langle T^{\mu}_{\mu} \right\rangle_{\rm eMIT}=4B-(2C)\mu^{2}_{B}$ with $C=\left[ 3/(2\pi^{2})\right] a_{2}$ for the effective MIT bag (eMIT) model \cite{Alford:2004pf}, being `$a_{2}$' the decisive parameter for positiveness or negativeness. {Notice that this last term is also related to the condensation term in the eMIT model.} 

Nevertheless, we need to verify if the presence of the `$C$' term is consistent or not with our CSS parametrization {for $P_{Q}(\epsilon)$} with $c^{2}_{s,Q}=1/\gamma=$ constant, i.e. {the speed of sound} being density independent, {in contrast to past more} realistic works (e.g. the eMIT or pQCD) consider density-dependent speeds of sound. 

After some manipulations (see Appendix E), the corresponding quark EoS, $P_{Q}=P_{Q}(\epsilon)$, becomes\footnote{We verified that this pressure reproduces the widespread EoSs of Refs. \cite{Zhang:2020jmb,VasquezFlores:2010eq,Pereira:2017rmp}. However, Ref. \cite{Zhang:2020jmb} does not justify the choice of positive sign, whereas Refs. \cite{VasquezFlores:2010eq,Pereira:2017rmp} selects correctly the negative sign, although without explicit explanation.} (assuming $N~$=constant implying a fixed number of degrees of freedom in the quark phase)
\begin{multline}
\label{eq:pressCSSdens}
P_{Q}=\frac{\epsilon}{\gamma}-\left(\frac{\gamma+1}{\gamma}\right)B+
\frac{C^{2}_{r}}{8N}\left(\frac{1-\gamma^{2}}{\gamma^{2}}\right)\\
\pm \frac{C_{r}}{8N}\left(\frac{\gamma+1}{\gamma}\right)\sqrt{\left(\frac{\gamma-1}{\gamma}\right)^{2} C^{2}_{r}+\frac{16N}{\gamma}(\epsilon-B)},
\end{multline}
where `$C_{r}$' is defined in Appendix E. The associated (squared) speed of sound is (still with $N$=constant)
\begin{multline}
\label{eq:cs2cond}
c^{2}_{s,Q}=\frac{dP}{d\epsilon}=\frac{1}{\gamma}\\
+\frac{C_{r}}{4N}\left(\frac{1-\gamma^{2}}{\gamma^{2}}\right)\left(\frac{dC_{r}}{d\epsilon}\right) \pm \left(\frac{\gamma+1}{\gamma}\right)\frac{1}{8N}\sqrt{(...)}\left(\frac{dC_{r}}{d\epsilon}\right) \\
 \pm \left(\frac{\gamma+1}{\gamma}\right)\frac{1}{\sqrt{(...)}}\left[\frac{C^{2}_{r}}{4N}\left(\frac{\gamma-1}{\gamma}\right)^{2} \left(\frac{dC_{r}}{d\epsilon}\right)+\frac{2C_{r}}{\gamma}\right],
\end{multline}
where `$\sqrt{(...)}$' means the square root in Eq. (\ref{eq:pressCSSdens}). Interestingly, this Eq. (\ref{eq:cs2cond}) indicates that constant `$C$' (leading to $C_{r}=C$) or running `$C$' (leading to $C_{r}\neq C$) should not be allowed in the EoS to keep our original assumption of $c^{2}_{s,Q}=1/\gamma$. Thus, at $\mu_{B} \neq 0$ (in marked contrast to finite isospin densities), one must put $C \equiv0$ if we give the speed of sound a constant value in a given range of baryon densities {within this CSS parametrization}.

With this in mind, Eq. (\ref{eq:traceCSSmu}) is reduced to (with $C =0$ and $N=~$constant)
\begin{equation}
\left\langle T^{\mu}_{\mu} \right\rangle_{\rm Q} = N(\gamma-3)\mu^{1+\gamma}_{B} +4B.
\end{equation}
It might be negative if $\gamma<3$, or equivalently, $c^{2}_{s, Q}>1/3$. In this sense, the negativeness of $\Delta$ comes directly from large $c^{2}_{s,Q}$ and not a condensation term (as in isospin matter) which spoils the consistency of the CSS parametrization. Of course, other quark models having control on the degrees of freedom and interactions can clarify this point.

\subsection{Insights for non-perturbative QCD physics}

\begin{figure}
  \vspace*{-0.55cm}	
  \hspace*{-0.6cm}
  \includegraphics[width=0.56\textwidth]{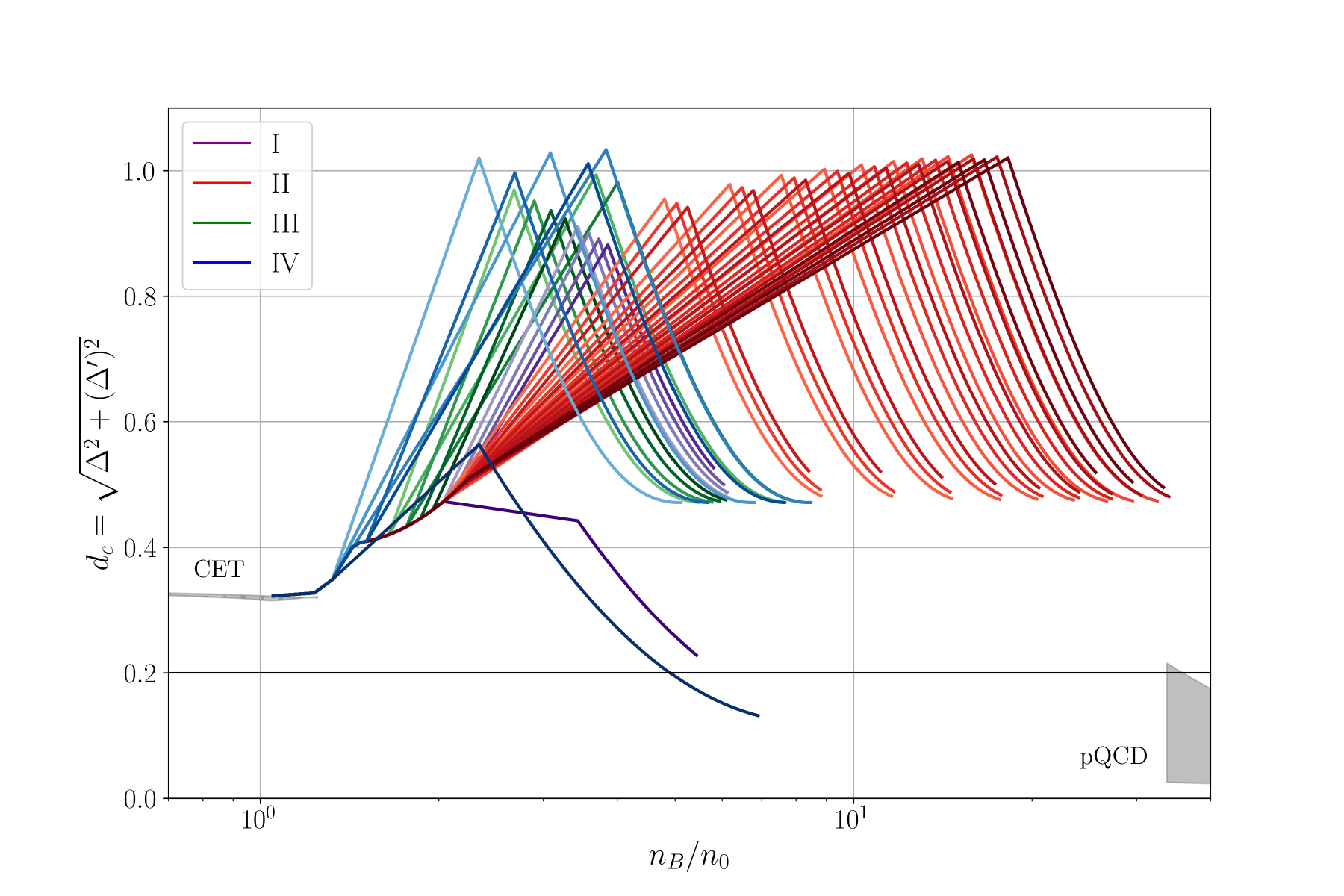}
  \caption{\label{fig:conform} Quantitative measure of conformality through $d_{c}= \sqrt{\Delta^2 + (\Delta')^2}$ \cite{Annala:2023cwx} as a function of the normalized baryon density, $n_{B}/n_{0}$. Notice that we keep the colors for each twin-star category as in previous figures. Besides, we draw the horizontal conformality transition at $d_{c}=0.2$. Again, the CET (thin) and pQCD (thick) grey bands are also showed.}
\end{figure}

Last year, in Ref. \cite{Annala:2023cwx}, it was postulated a new parameter $d_{c}\equiv \sqrt{\Delta^{2}+(\Delta')^{2}}$ (being $\Delta'=d\Delta/d\ln \epsilon$) as a quantitative (and non-perturbative) measure of {\it conformality} in strongly interacting matter. In particular, they found \cite{Annala:2023cwx} $\Delta'=1/3-\Delta$ for first-order phase transitions. Besides, they established that near conformal (deconfined) matter appears when $d_{c}<0.2$, thus preventing first-order transitions from masquerading as conformalized matter (see Ref. \cite{Annala:2023cwx} for details).  

We show in Fig. \ref{fig:conform} results for our four ({\it rapid}) twin-star categories. Compared to Fig.~1 of Ref. \cite{Annala:2023cwx}, one can extract several new features. First of all, the magnitudes for `$d_{c}$' {increase sizably}, i.e. now even reaching values of around 1 due to the jumps in $\Delta n_{B}$. This `high' values were considered impossible in the computations of Ref. \cite{Annala:2023cwx} which maximally reached $d_{c}\sim 0.6$ taking into account only smooth phase transitions, e.g. rapid crossovers. Secondly, our results at the critical point ({which} tend to decrease somewhat quickly) never go below the conformal transition point of $d_{c}=0.2$, thus potentially {being classified} as non-conformal (except with a pair of cases from categories I and IV). Thirdly, unlike in Ref. \cite{Annala:2023cwx}, our $d_c$'s could approach the conformal pQCD regime from above, i.e. $d_{c}>0.2$, thus excluding any possible existence of conformal matter at intermediate densities even having stiff QM, that is to say, ultra-dense quark matter could still be non-conformal, being purely conformal at unrealistic densities, {such as} $n_{B}\geq 40n_{0}$.

We stress that the findings in Fig.~\ref{fig:conform} in this work are not the only ones in disagreement with those of Ref. \cite{Annala:2023cwx}. Other works that appeared along the last year performing more general (mostly Bayesian) investigations of this conformality parameter `$d_{c}$' sometimes strongly disagree with findings of Ref. \cite{Annala:2023cwx}, where it was found the range of values between $0.1 \lesssim d_{c}\lesssim 0.45$. In particular, {some of these works} pointed out \cite{Takatsy:2023xzf,Malik:2024qjw,Albino:2024ymc,Marquez:2024bzj} (through studies with nucleonic or hybrid models) that `$d_{c}$' should be carefully interpreted. For instance, some of these robust explorations (e.g. Ref. \cite{Marquez:2024bzj} within the so-called extended Nambu-Jona-Lasinio model) always found $d_{c}>0.2$ with/without an adjustment to converge to pQCD at high densities. Unfortunately, none of these works considered the presence of strong transitions possibly giving $d_{c}\sim 1$, as in the present work.

\begin{figure}	
  \includegraphics[width=0.48\textwidth]{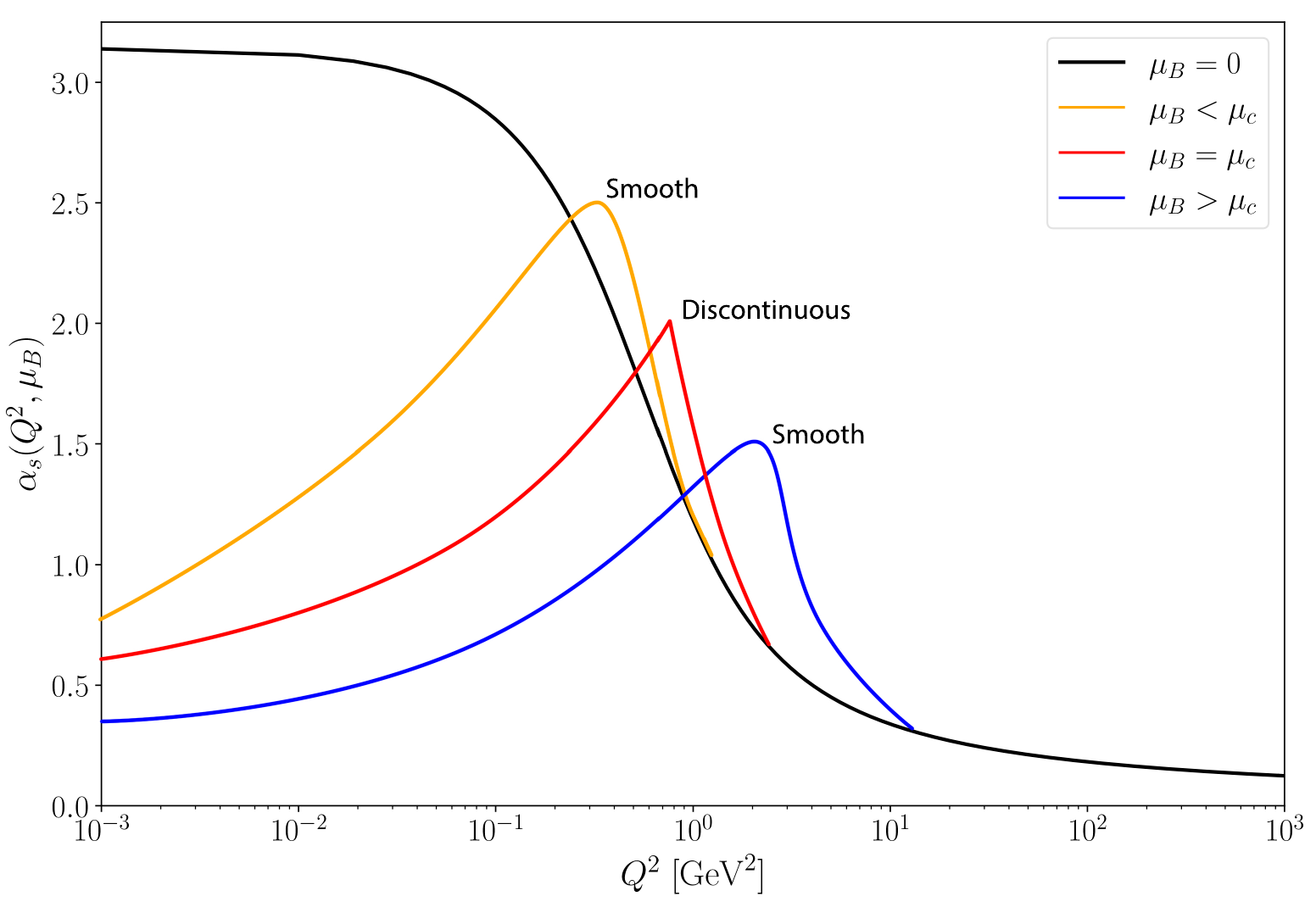}
  \caption{\label{fig:strongQCD} Schematic representation for the behavior of the QCD strong coupling in a dense medium, $\alpha_{s}(Q^2,\mu_B)$ (also depending on the energy-momentum scale $Q^2$) inferred from $\Delta$. The black curve for the vacuum ($\mu_B \neq 0$) case was obtained from Eq. (1) of Ref. \cite{Deur:2023dzc}. The other curves represent our hypothesized behavior for `$\alpha_s$' before ($\mu_{B}<\mu_c$), at ($\mu_{B}=\mu_c$), and after ($\mu_{B}>\mu_c$) the first-order phase transition occurs.}
\end{figure}

Finally, {inspired by all these findings related to the} discontinuities present in $\Delta=\Delta(\epsilon)$, $\Delta=\Delta(\mu_{B})$, $d_{c}=d_{c}(n_{B}/n_{0})$ around the transition point, {we are lead to hypothesize the possible behavior of the QCD strong coupling $\alpha_{s}$ since} until now no full and reliable QCD solution is known. {Thus, by looking at} Eq. (\ref{eq:trace}), one can see the strong connection of $\Delta$ with $\beta_{\rm QCD}$, or more specifically, to $\alpha_{s}$ (assuming the chiral condensates contribute more or less as constants). In this sense, it is reasonable to infer that the discontinuities present in the dense trace anomaly come from {the} discontinuities in $\alpha_{s}$ around the critical baryochemical potential. 

{We sketch in Fig. \ref{fig:strongQCD} our proposal for $\alpha_{s}$, where one can see (note the horizontal log scale) a general decreasing trend at deep infrared energies, $Q^{2}$. In particular, for $\mu_{B} < \mu_{c}$, $\alpha_{s}$ has an smooth maximum-like point implying $\beta_{\rm QCD}=0$ or $\Delta=0$, i.e. conformality could exist at intermediate densities, as in Ref.~\cite{Fujimoto:2022ohj}. More importantly, $\alpha_{s}$ induces an smooth behavior of $\Delta$ for $\mu_{B} < \mu_{c}$. However, at $\mu_{B}=\mu_{c}$, $\alpha_{s}$ displays a fully discontinuous behavior with the presence of a peak giving ($\partial \alpha_{s}/\partial \mu_{B})(\mu^{+}_B \to \mu_{c})\neq (\partial \alpha_{s}/\partial \mu_{B})(\mu^{-}_B \to \mu_{c})$ (see also Ref. \cite{Arefeva:2024poq}), i.e. the derivative of $\alpha_{s}$ is different when going from the left than the right.  This means that $\beta_{\rm QCD}$ and $\Delta$ are discontinuous at the transition point, as obtained in our calculations in sections above through twin stars. Then, for $\mu_{B}>\mu_{c}$, $\alpha_{s}$ is again smooth (as $\Delta$) with a maximum. It should be realized that these three maxima are displaced to the right for increasing $\mu_{B}$. Notice also that we put in Fig. \ref{fig:strongQCD} the results for $\alpha_{s}$ in vacuum for comparison \cite{Deur:2023dzc}. It is worth to mention that the distinct values of $\alpha_{s}$ around $Q^{2} \sim 10^{-3}~{\rm GeV^{2}}$ depend upon the infrared scenarios, e.g. the {\it decoupling} and {\it scaling} \cite{Deur:2016tte,Deur:2023dzc}.}

We stress that similar (continuous) findings were obtained from effective models in the past (see, e.g. Refs. \cite{Steffens:2004sg,Braun:2006jd,Fu:2019hdw}) at finite temperatures. At $\mu_{B}\neq 0$, we are not aware of similar non-perturbative studies for which we believe it is necessary a detailed study. For instance, {only} studies in vacuum within the functional-renormalization group, Schwinger-Dyson and machine-learning techniques (based upon experimental data) \cite{Deur:2023dzc,Wang:2023poi} of $\alpha_{s}$ for $Q^{2}\ll 1$ are known. Of course, our hypothesis could only be verified by future LQCD calculations at finite baryon densities (e.g. within quantum computing approaches \cite{Yamamoto:2021fjs}) find first-order transition features in, e.g., $\alpha_{s}(Q^{2},\mu_{B})$, as in hot LQCD \cite{Bazavov:2009zn,Borsanyi:2013bia}.

\section{Conclusions and outlook}
\label{sec:conclusion}

We performed a comprehensive analysis of the dense QCD matter (normalized) trace anomaly, $\Delta$, at the interior of twin NSs (characterized by having sizeable cores of QM). Within the four categories of twin stars explored with rapid and slow {junction} conditions in the radial pulsation equations, we found that their $\Delta$ show an abrupt decreasing trend to negative values at intermediate densities after the onset to QM. This is in contrast {to the findings of LQCD giving a $\Delta\geq 0$ at all} temperatures \cite{Bazavov:2009zn,Borsanyi:2013bia}. {However, our results are in agreement with LQCD studies for cold isospin matter} \cite{Abbott:2023coj}. {This negativity of $\Delta$ could also be understood as a requirement for twin stars exist in nature.} This is due to large values speed of sound that twin stars require, i.e. stiffer QM EoS. 

In particular, we were only able to find two stellar stable configurations with $c^{2}_{s, Q} = 0.5$, one for Category I and another for Category IV, while the remaining families require $c^{2}_{s, Q}=1$, in the rapid scenario.  All these differences with past studies come from the presence of a strong discontinuous phase transition representing the QCD deconfinement transition. It is worth mentioning that our results go into the opposite direction of Ref.~\cite{Fujimoto:2022ohj}, where they conjecture $\Delta\geq 0$ at all NS densities. Interestingly, our findings slightly agree with Ref. \cite{Ecker:2022dlg}, where they allowed $\Delta$ to be generically negative at NS cores, but not having too negative values while smoothly/slowly approaching conformality. Another fundamental difference with this and other current works comes from noticing that the decreasing behavior of $\Delta$ in twin stars is steeply abrupt, i.e. large negative slopes, which directly depend on the stiffness of the QM EoS, i.e. on the value of the (squared) speed of sound, in other words, higher values of $c^{2}_{s, Q}$ leading to a steeper negative slopes {of $\Delta$}.

\begin{figure*}
\vspace*{-0.55cm}
\hspace*{-0.71cm}
\includegraphics[width=0.54\textwidth]{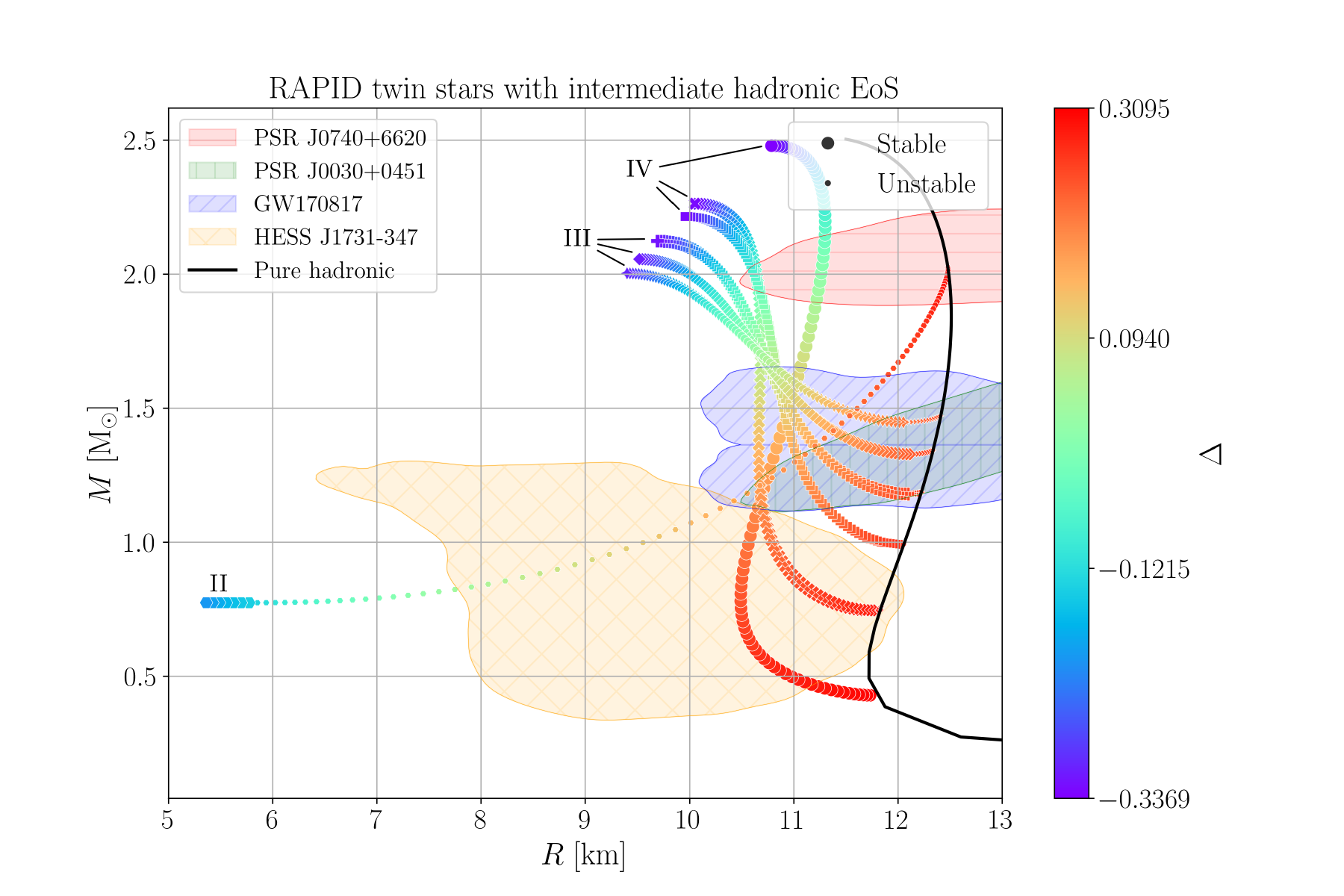}
\hspace*{-1.1cm}
\includegraphics[width=0.54\textwidth]{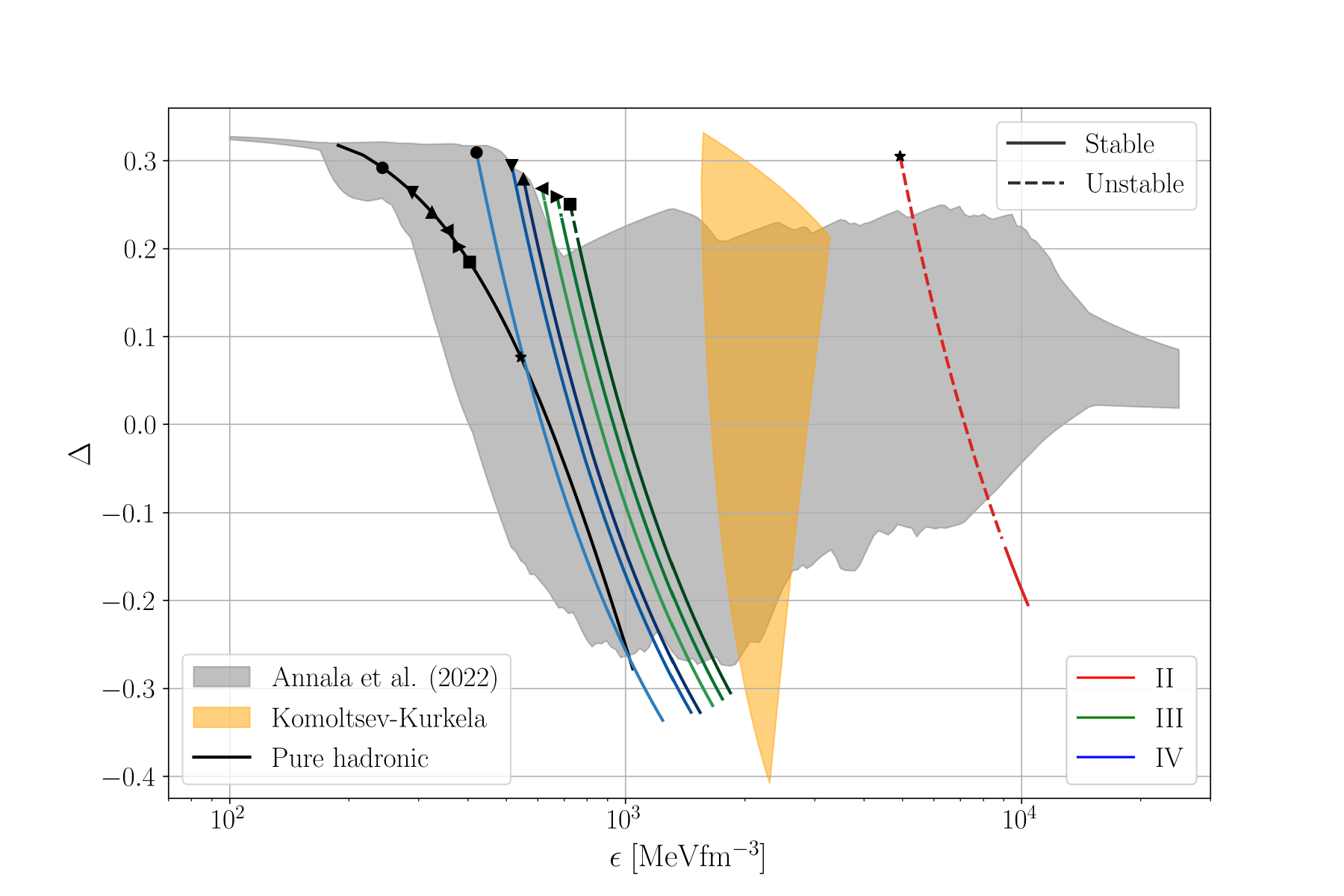}
\caption{\label{fig:intermediate} Same captions and constraints as in Fig. \ref{fig:catI_II} but for all rapid twin stars employing the intermediate CET EoS in the hadronic sector. Note that one gets again the sudden decrease of $\Delta$ although still in agreement with the band of Annala et al. (2022) \cite{Annala:2021gom} except the rightmost curve being marginally stable (see also the leftmost $M$--$R$ relation in the left panel).}
\end{figure*}

We also found that the presence of strong transitions modify highly non-trivially expectations for the behavior of the normalized pressures $P/P_{\rm SB}$ and the conformal factor `$d_{c}$' (being tightly related to $\Delta$), in particular, large values for both, $P/P_{\rm SB}>1$ and $d_{c}>0.6$. Besides, the well-known band of Komoltsev-Kurkela \cite{Komoltsev:2023zor} was proven to be insufficient when applied to twin stars {since it is maximally valid at} $n_{B}=10n_{0}$ {to ensure convergence} to pQCD at $40n_{B}$. {However, our calculations indicate that the cores of twin stars have larger values around $10\leq n_{B}/n_{0} \lesssim 30$. In this sense,} we believe that more delicate studies of these {kind of constraints, e.g. the band of Komoltsev-Kurkela,} must be redone explicitly including strong transitions by, e.g. employing recent parametric representations of the NS EoS allowing strong discontinuities coming from phase transitions \cite{Lindblom:2024dys}. Another similar research direction could extend {current} Bayesian studies to include sequential phase transitions, i.e. two strong discontinuities \cite{Alford:2017qgh}, which from all our figures above seem to be a reasonable mechanism to approach pQCD, even having large $P/P_{\rm SB}$ and `$d_{c}$'. 

{Moreover,} although there are some recent works (see, e.g. Ref. \cite{Naseri:2024rby}) stating that twin stars are rare in nature, there are still several microphysics unknowns relavant to build a fully appropiate twin star EoS. For instance, this last study of Ref. \cite{Naseri:2024rby} within a fully dynamical general-relativistic framework does not take into account at any moment the phase conversion dynamics between hadronic matter and QM, {which eventually leads} to the rapid/slow {phase-conversion junction conditions} \cite{Pereira:2017rmp} in the quasi-stationary limit.

On the other hand, we hypothesized that the existence of discontinuities in $\Delta$ come from fundamental discontinuities in $\beta_{\rm QCD}$, which in turn are manifested in $\alpha_{s}(Q^{2},\mu_{B})$, where non-perturbative QCD physics could be manifested through these discontinuities. Currently, estimates of this coupling are {only} obtained in vacuum, inferring a smooth behavior in the infrared limit \cite{Deur:2023dzc,Wang:2023poi}.

We leave as future work detailed investigations concerning different thermodynamic constructions \cite{Glendenning:2000} at the phase-transition point,  non-equilibrium effects in the junction conditions \cite{Rau:2023wiq}, anisotropy modifications \cite{Horvat:2010xf}. Besides, it would be worth to investigate the effects of the recent proposal of Ref. \cite{Marczenko:2024uit} for an averaged (squared) speed of sound, generalizing the Seidov's criterion.

\begin{acknowledgments}
The authors thank Tyler Gorda and Ryan Abbott for sharing data of Refs. \cite{Gorda:2022lsk} and \cite{Abbott:2023coj}, respectively. This work was partially supported by INCT-FNA (Process No. 464898/2014-5). V.P.G. and L.L. acknowledges support from CNPq, CAPES (Finance Code 001), and FAPERGS. V.P.G. was partially supported by the CAS President's International Fellowship Initiative (Grant No.  2021VMA0019). J.C.J. is supported by Conselho Nacional de Desenvolvimento Cient\'{\i}fico e Tecnol\'ogico (CNPq) with Grant No. 151390/2024-0.
\end{acknowledgments}
\section*{Appendix A: \\TRACE ANOMALY OF \\HOLOGRAPHIC QUARK MATTER}
Some years ago, the authors of Ref. \cite{Hoyos:2016zke} found a {novel advance towards obtaining} the trace anomaly of dense matter {in the nonperturbative sector of QCD} applying the AdS/CFT duality. They obtained
\begin{equation}
\epsilon (P)=3P+\frac{\sqrt{3}m^{2}}{2\pi}\sqrt{P},
\end{equation}
with $m\approx 308.55$ MeV {(see Ref. \cite{Hoyos:2016zke} for details)}. Unfortunately, the way this equation was given is not very useful for our purposes{, i.e. $P=P(\epsilon)$ instead of $\epsilon=\epsilon(P)$}. Thus, we manipulate it algebraically, obtaining
\begin{equation*}
P=\frac{\epsilon}{3}+\frac{m^{4}}{24\pi^{2}}\pm \frac{\sqrt{3}m^{2}}{36\pi}\sqrt{12\epsilon+\frac{3m^{4}}{4\pi^{2}}}=P(\epsilon).
\end{equation*}
The ambiguity of the signs is solved by making this result consistent with the original one, i.e.
\begin{equation*}
\epsilon-3P=\frac{\sqrt{3}m^{2}}{2\pi}\sqrt{P}=-\frac{m^{4}}{8\pi^2}\mp \frac{\sqrt{3}m^{2}}{12\pi}\sqrt{12\epsilon+\frac{3m^{4}}{4\pi^{2}}}>0,
\end{equation*}
implying the need of a negative sign in $P=P(\epsilon)$ above. It is also worth to remark that this choice agrees with one boundary condition of the TOV equations, i.e.  $P=0$ at a given energy density related to a stellar surface. A positive sign would spoil this condition, {producing stars with unrealistically large radii.}

\begin{figure}[!t]
  \vspace*{-0.55cm}	
  \hspace*{-0.6cm}
\includegraphics[width=0.56\textwidth]{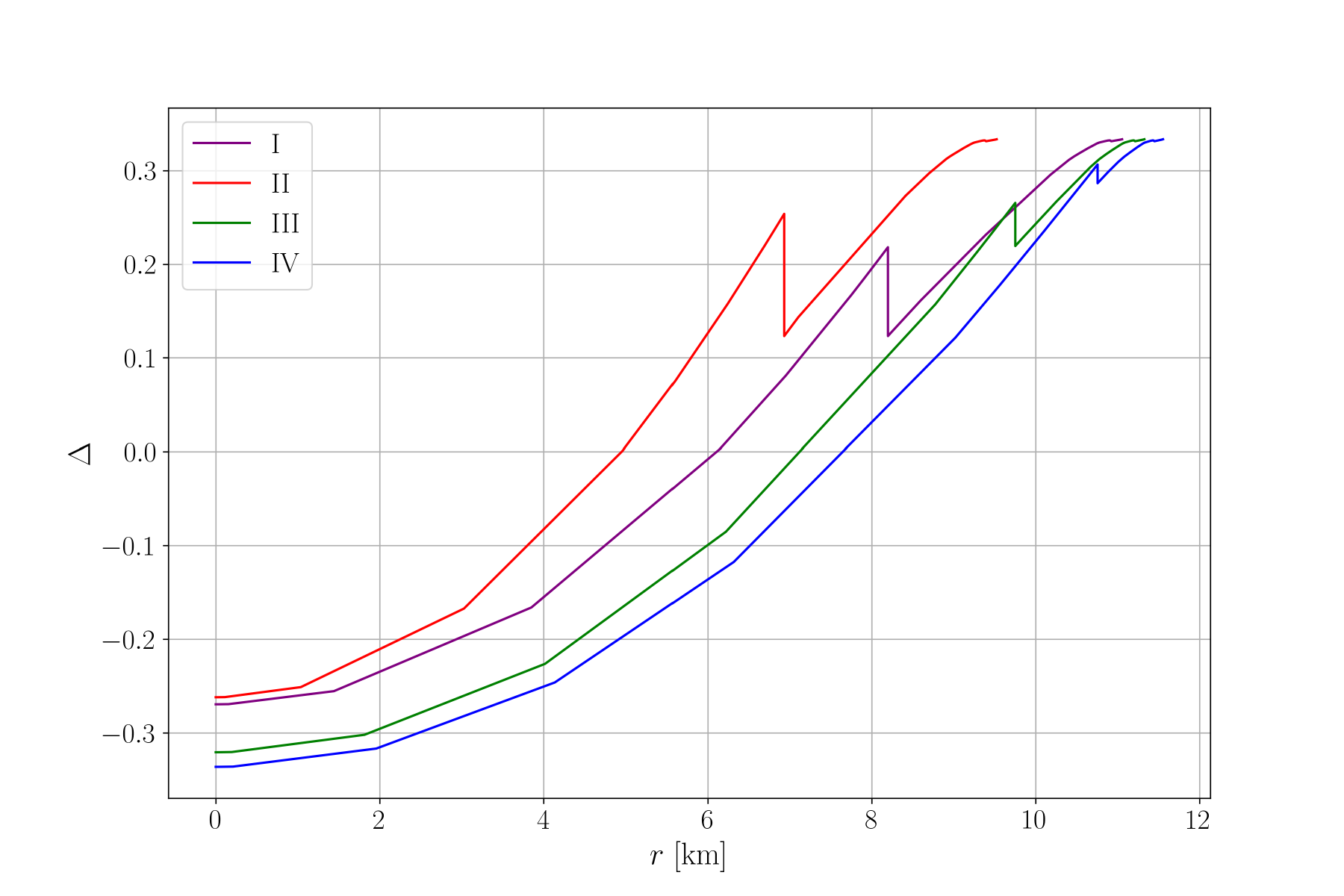}
\caption{Radial profiles of the dense trace anomaly for maximally-stable rapid twin stars obtained from the parameters listed in Table~\ref{tab:css_parameters}. Notice the pronounced peaks at fixed radii for each category characterizing the phase-transition point. These curves were obtained from $\Delta(r)=1/3-P_{\rm max}(r)/\epsilon_{\rm max}(r)$, where $\left\lbrace P_{\rm max}, \epsilon_{\rm max}\right\rbrace$ are solutions of the TOV equations for the maximal-mass twin stars.}
\label{fig:stiff_radial_profile}
\end{figure}

\section*{Appendix B: \\RESULTS FOR TWIN-STAR TRACE ANOMALIES WITH INTERMEDIATE CET EOS}
It is worth to test if our main conclusions for the {normalized} twin star trace anomaly, $\Delta$, are still true when using the intermediate CET EoS of Ref. \cite{Hebeler:2013nza} with the following agnostic GPP parametrization: $\log_{10}[\rho_0 /({\rm g~cm^{-3}})] = 13.865$, $\log_{10} K_1 = -27.22$, $\Gamma_1 = 2.748$, $\Gamma_2 = 6.5$ and $\Gamma_3 = 3.25$. Since this hadronic EoS is not as stiff as the one employed along this work, we found a significant smaller number of twin-star EoSs that satisfy the two-solar mass constraint. Thus, we found that twin stars in Category I are not possible to exist due to the lack of stiffness in the intermediate CET hadronic EoS. On the other hand, we present in Fig. \ref{fig:intermediate} our predictions for all remaining categories with rapid-conversion junction conditions (where the usual stability criterion, $\partial M/\partial \epsilon_{c}\geq 0$ works) for the $M$--$R$ relation (left panel) and the {normalized} trace anomaly as function of energy density (right panel), following the same color scheme as previous {related} figures in the main text. Results for slow junction conditions are consistent to those presented in the main text, i.e. only Category II configurations present a larger number of valid EoSs when compared to the rapid scenario.

\section*{Appendix C: \\RADIAL PROFILES FOR \\THE TWIN-STAR TRACE ANOMALIES}
Apart from our main results, one can also be interested in the values reached by $\Delta$ at the core of twin stars. In order to answer that question, we show in Fig.~\ref{fig:stiff_radial_profile} the radial profiles for the twin star trace anomalies for the same of four EoSs (one for each category) of Table~\ref{tab:css_parameters} considering only rapid conversions. In particular, we chose to display only our findings of the {normalized} trace anomaly $\Delta$ inside the maximum-mass twin stars for simplicity. This configuration is also the last one with a positive fundamental eingenfrequency, i.e. the last stable one. Finally, we verified that other rapid-conversion (as well as slow-conversion) twin stars show basically the same behavior as the maximally-stable twin star.

\section*{Appendix D: \\ROBUSTNESS OF $\Delta <0$ \\FOR THE POLYTROPIC MODEL \\OF NUCLEAR MATTER}
As mentioned in Sec. \ref{sec:disc}, we {pass to} prove that even {by replacing the quantum hadrodynamic model by} a polytropic one for nuclear matter, one still would get $\Delta<0$ at intermediate densities.

For a given polytrope `$i$' within a piecewise polytropic pressure, one has that 
\begin{equation*}
P_{\rm poly}=\kappa_{i}{n}^{\Gamma_{i}}, ~\epsilon_{\rm poly}=\frac{\kappa_{i}n^{\Gamma_{i}}}{(\Gamma_{i}-1)}+m_{0}n,
\end{equation*}
for the pressure and energy density, respectively, with `$m_{0}$' the particle's mass, `$n$' particle's density, and $\Gamma$ as the polytropic index. This result was obtained by using $P=n^{2}d(\epsilon/n)/dn$. Thus, the {corresponding} (unnormalized) polytropic dense trace anomaly becomes
\begin{equation*}
\left\langle T^{\mu}_{\mu}\right\rangle_{\rm poly}=\epsilon_{\rm poly}+3(\epsilon_{\rm poly}-m_{0}n)(1-\Gamma_{i}).
\end{equation*}
Interestingly, {one can} notice that in the limit of $\epsilon_{\rm poly}\gg m_{0}n$, i.e. somewhat high densities {at the nuclear scale} or equivalently intermediate QCD densities, one gets
\begin{equation*}
\left\langle T^{\mu}_{\mu} \right\rangle_{\rm poly}\simeq\epsilon_{\rm poly}\left(1+3(1-\Gamma_{i})\right)=3\epsilon_{\rm poly}\left(\frac{4}{3}-\Gamma_{i}\right),
\end{equation*}
or also equivalently for the normalized case
\begin{equation}
\Delta_{\rm poly}=\frac{\left\langle T^{\mu}_{\mu} \right\rangle_{\rm poly}}{3\epsilon_{\rm poly}}=\frac{4}{3}-\Gamma_{i}.
\end{equation}
In other words, $\left\langle T^{\mu}_{\mu} \right\rangle_{\rm poly}<0$ (or $\Delta_{\rm poly}<0$) when $\Gamma_{i}>4/3$. In other words, all works in the literature (see e.g. Refs. \cite{Hebeler:2013nza,Annala:2019puf} surpassing $\Gamma=4/3$) employing polytropes at intermediate densities `$n$' will {implicitly be probing} negative $\left\langle T^{\mu}_{\mu} \right\rangle_{\rm poly}$. Besides, {since} at intermediate densities $P\approx \epsilon$ {(since $c^{2}_{s}\to 1$ in NS matter \cite{Annala:2019puf})}, then the corresponding polytropic speed of sound behaves as
\begin{equation*}
c^{2}_{s}=\frac{\Gamma_{i} P}{\epsilon + P}\approx \frac{\Gamma_{i}}{2}.
\end{equation*}
Thus, $\Gamma_{i}>4/3$ implies $c^{2}_{s}>2/3\approx0.67$, i.e. $\left\langle T^{\mu}_{\mu} \right\rangle_{\rm poly}<0$ occurs for large squared speeds of sound, e.g., $>0.5$.

Of course, one can make $\left\langle T^{\mu}_{\mu} \right\rangle_{\rm poly}$ positive again by adding an ultra-dense EoS, e.g. the bag model (or pQCD like in Ref. \cite{Annala:2019puf}) for which $\left\langle T^{\mu}_{\mu} \right\rangle_{\rm bag}=4B$ (`$B$' being the MIT bag constant). {In total,} one would get
\begin{equation*}
\left\langle T^{\mu}_{\mu} \right\rangle_{\rm poly}+\left\langle T^{\mu}_{\mu} \right\rangle_{\rm MIT}=3\epsilon_{\rm poly}\left(\frac{4}{3}-\Gamma_{i}\right)+4B,
\end{equation*}
which can make the total trace anomaly to be positive.

\section*{Appendix E: \\PROOFS ABOUT THE CONSTANT-SPEED-OF-SOUND MODEL}
We pass to prove Eq. (\ref{eq:pressCSSdens}) from Eq. (\ref{eq:new}). First, we obtain the corresponding energy density using known thermodynamic relations
\begin{multline*}
\epsilon_{Q}=-P_{Q}+\mu_{B}n^{Q}_{B}=\\
N(\mu_{B})\gamma\mu^{1+\gamma}_{B}+\left(\frac{\gamma-1}{2}\right)C(\mu_{B})\mu^{(1+\gamma)/2}_{B}+\\
\left(\frac{\partial N}{\partial \ln \mu_{B}}\right)\mu^{1+\gamma}_{B}+\left(\frac{\partial C}{\partial \ln \mu_{B}}\right)\mu^{(1+\gamma)/2}_{B}+B,
\end{multline*}
being appropriately rewritten as (generically with $N=N(\mu_{B})$ and $C=C(\mu_{B})$)
\begin{multline*}
\left(N\gamma+\frac{\partial N}{\partial \ln \mu_{B}}\right) \mu^{1+\gamma}_{B}+\\
\left[\left(\frac{\gamma-1}{2}\right)C+\frac{\partial C}{\partial \ln \mu_{B}}\right]\mu^{(1+\gamma)/2}_{B}-(\epsilon_{Q}-B)=0,
\end{multline*}
where we pass to define $N_{r}\equiv N+(1/\gamma)({\partial N}/{\partial \ln \mu_{B}})$ and $C_{r}\equiv C+ [2/(\gamma-1)]({\partial C}/{\partial \ln \mu_{B}})$ as {\it running} generalizations. This {last} equation can be identified with a 2nd-order algebraic one with $x\equiv \mu^{(1+\gamma)/2}_{B}$, thus giving two real roots given by
\begin{multline*}
\mu^{(1+\gamma)/2}_{B}=-\left(\frac{\gamma+1}{\gamma}\right)\frac{C_{r}}{4N_{r}}\\
\pm \left(\frac{C_r}{4N_{r}}\right)\sqrt{\left(\frac{\gamma-1}{\gamma}\right)^{2}+\frac{16N_{r}}{\gamma C^{2}_{r}}(\epsilon-B)}. 
\end{multline*}
This last result must be introduced in Eq. (\ref{eq:new}), reordered and simplified to finally give Eq. (\ref{eq:pressCSSdens}). For the same reasons discussed in Appendix A and footnote {9}, we chose the negative sign in the above solutions for `$x$'.

On the other hand, leaving aside the CSS parametrization with $c^{2}_{s}=\text{constant}=1/\gamma$, one interesting effect happens within our results above {if we} consider that `$\gamma$' and $c^{2}_{s}$ are independent quantities, i.e. $c^{2}_{s, Q} \neq 1/\gamma$, with $c^{2}_{s}=dP/d\epsilon$. Thus, by fixing $N=~$constant, `$C$' running and $\gamma=1$, we get
\begin{equation*}
\mu_{B}=\sqrt{\frac{\epsilon-B}{N}}~{\rm or}~\ln\mu_{B}=(1/2)\ln[(\epsilon-B)/N],
\end{equation*}
which after being introduced in Eq. (\ref{eq:pressCSSdens}) gives
\begin{multline*}
P_{Q}=\epsilon_{Q}-2B \pm \sqrt{\frac{\epsilon_{Q}-B}{N}}\left(C+\lim_{\gamma \to 1}\frac{1}{\gamma-1}\frac{\partial C}{\partial \ln \mu_{B}}\right)
\end{multline*}
or more appropriately
\begin{multline*}
P_{Q}=\epsilon_{Q}-2B \\
\pm \sqrt{\frac{\epsilon_{Q}-B}{N}}\left(C+\lim_{\gamma \to 1}\frac{2}{\gamma-1}\frac{\partial C}{\partial \ln (\epsilon_{Q}/N)}\right).
\end{multline*}
Thus, assuming a running non-zero condensation term makes to diverge the EoS as $\gamma \to 1$ (also its $c^{2}_{s}$) coming from a fast increment in the running behavior of `$C$'. If one assumes it is constant, then the appropiate choice of sign would be minus in order to not surpass the causality limit ($c^{2}_{s}=1$) at high energy densities. In any case, something is evident for a generic $\gamma<1$: large values for the speeds of sound at intermediate densities are related to strong effects from the non-perturbative condensation term, which commonly only enters as subdominant terms, for instance, in CFL quark matter \cite{Alford:2004pf}. 


\end{document}